%% file: paper.tex
\documentclass[usenatbib]{mn2e}
\usepackage{amsmath}
\usepackage{natbib}
\usepackage{epsfig}
\usepackage{txfonts}

\input{Befehle}

\newcommand{\ion}[2]{{\text{{\sc #1}\,{\sc #2}}}}
\newcommand{\omegab}{{\Omega_{\rm b}h^2}}

\newcommand{\YHe}{Y_{\rm p}}
\newcommand{\Nnu}{N_{\nu}}
\newcommand{\Planck}{{\sc Planck }}

\usepackage{color}
\newcommand{\changeJ}[1]{#1}
\newcommand{\changeI}[1]{{#1}}

\usepackage{hyperref}
\usepackage{grffile}
\usepackage{graphics}
\usepackage{booktabs}

\title[Precise cosmological parameter estimation using {\sc CosmoRec}]
{Precise cosmological parameter estimation using {\sc CosmoRec}}

\author[Shaw and Chluba]{J.~R.~Shaw$^{1}$\thanks{E-mail:
   jrs65@cita.utoronto.ca} and J.~Chluba$^{1}$\thanks{E-mail:
   jchluba@cita.utoronto.ca}
  \\
$^{1}$ Canadian Institute for Theoretical Astrophysics, 60 St. George Street,
Toronto, ON M5S 3H8, Canada
}

\begin{document}

\date{Received 2011; Accepted 2011}

\maketitle

\begin{abstract}
\changeJ{We} use the new cosmological recombination code, {\sc
    CosmoRec}, for parameter estimation in the context of (future)
  precise measurements of the CMB temperature and polarization
  anisotropies.
  We address the question of how previously neglected physical
  processes in the recombination model of {\sc Recfast} affect the
  determination of key cosmological parameters, for the first time
  performing a {\it model-by-model} computation of the recombination
  problem.
  In particular we ask how the biases depend on different combinations
  of parameters, e.g. when varying the helium abundance or the
  effective number of neutrino species in addition {to} the standard
  six parameters.
  We also forecast how important the recombination corrections are for
  a combined {\Planck}, {\sc ACTPol} and {\sc SPTpol} data
  analysis.
  Furthermore, we ask which recombination corrections are really
  crucial for CMB parameter estimation, and whether an approach based
  on a redshift-dependent {\it correction function} to {\sc Recfast}
  is sufficient in this context.
%
%\changeJ{With the analysis carried out here we demonstrate that {\sc CosmoRec} 
%is ready for precise parameter estimation using {\Planck} data and beyond.}

\end{abstract}
%------------------------------------------------------------

\begin{keywords}
Cosmic Microwave Background: cosmological recombination, temperature anisotropies, radiative transfer
\end{keywords}

\section{Introduction}
\label{sec:Intro}
The {\Planck} Surveyor\footnote{\url{http://www.rssd.esa.int/Planck}} is
currently observing the temperature and polarization anisotropies of
the cosmic microwave background (CMB) with unprecedented accuracy,
constantly producing new exciting results.
By now it has successfully completed two full sky scans, and is about
to finished its third. On January 11th, 2011, the first release of
{\Planck} \changeJ{data} became available to the public, e.g. the Early
Release Compact Source Catalog \citep{AdeERCSC}, \changeJ{which also contains} the Early
Sunyaev-Zel'dovich Cluster Sample \citep{AdeESZCS}.
Using \changeI{{\Planck}} data, cosmologists will be able to
determine the key cosmological parameters with extremely high
precision, making it possible to distinguish between {various models}
of {\it inflation} \citep[e.g. see][]{Komatsu2010} by measuring the
precise value of the spectral index of scalar perturbations, $n_{\rm
  S}$, and constraining its possible running, $n_{\rm run}$.
In the near future {\sc
  SPTpol}\footnote{\url{http://pole.uchicago.edu/}} \citep{SPTpol}and
{\sc ACTPol}\footnote{\url{http://www.physics.princeton.edu/act/}}
\citep{ACTPol} will provide additional small scale $E$-mode
polarization data, complementing the polarization power spectra
obtained with {\Planck}.

These encouraging observational prospects have motivated various
independent groups \citep[e.g. see][]{Dubrovich2005, Chluba2006,
  Kholu2006, Switzer2007I, Wong2007, Jose2008, Karshenboim2008,
  Hirata2008, Chluba2008a, Jentschura2009, Labzowsky2009, Grin2009,
  Yacine2010} to assess how uncertainties in the theoretical treatment
of the {cosmological} recombination process could affect the science return of {\sc
  Planck} and future CMB experiments.
It was shown \citep{Jose2010} that {for the standard six parameter cosmology} in particular our ability to
measure the precise value of $n_{\rm S}$ and the baryon content of our
Universe will be compromised if modifications to the recombination
model of {\sc Recfast} \citep{SeagerRecfast1999, Seager2000} are
neglected.

These efforts are now coming to an end, and currently it appears that
{\it all} important corrections to the standard recombination scenario
have been identified \citep[e.g. see][for overview]{Fendt2009,
  Jose2010}.
This has lead to the development of two {{\it new} independent}
recombination codes, {\sc CosmoRec} \citep{Chluba2010b} and {\sc
  HyRec} \citet{Yacine2010c}, both of which include all important
corrections to the recombination problem, superseding the physical
model of {\sc Recfast}.
{\sc CosmoRec} and {\sc HyRec} allow fast and accurate computations of
the ionization history on a model-by-model basis, without the
necessity of {\it fudging}. \changeJ{A} detailed code-comparison is
currently in preparation, \changeJ{however, `out-of-the-box'
  comparisons indicate agreement at a level of $\sim 0.1\%-0.2\%$
  during hydrogen recombination ($z\sim 1100$).}

\begin{figure*}
\centering
\includegraphics[width=\columnwidth]{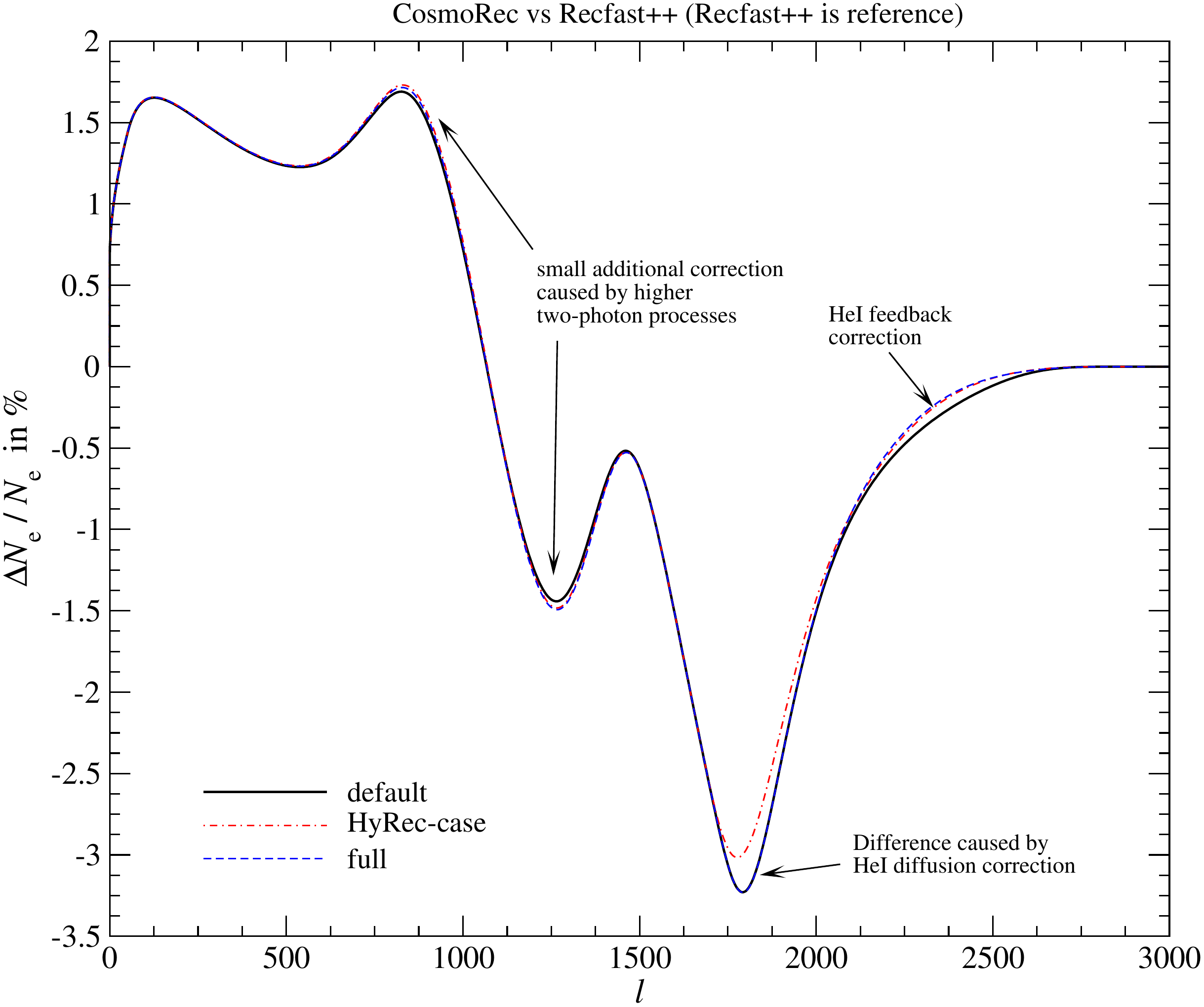}
\includegraphics[width=\columnwidth]{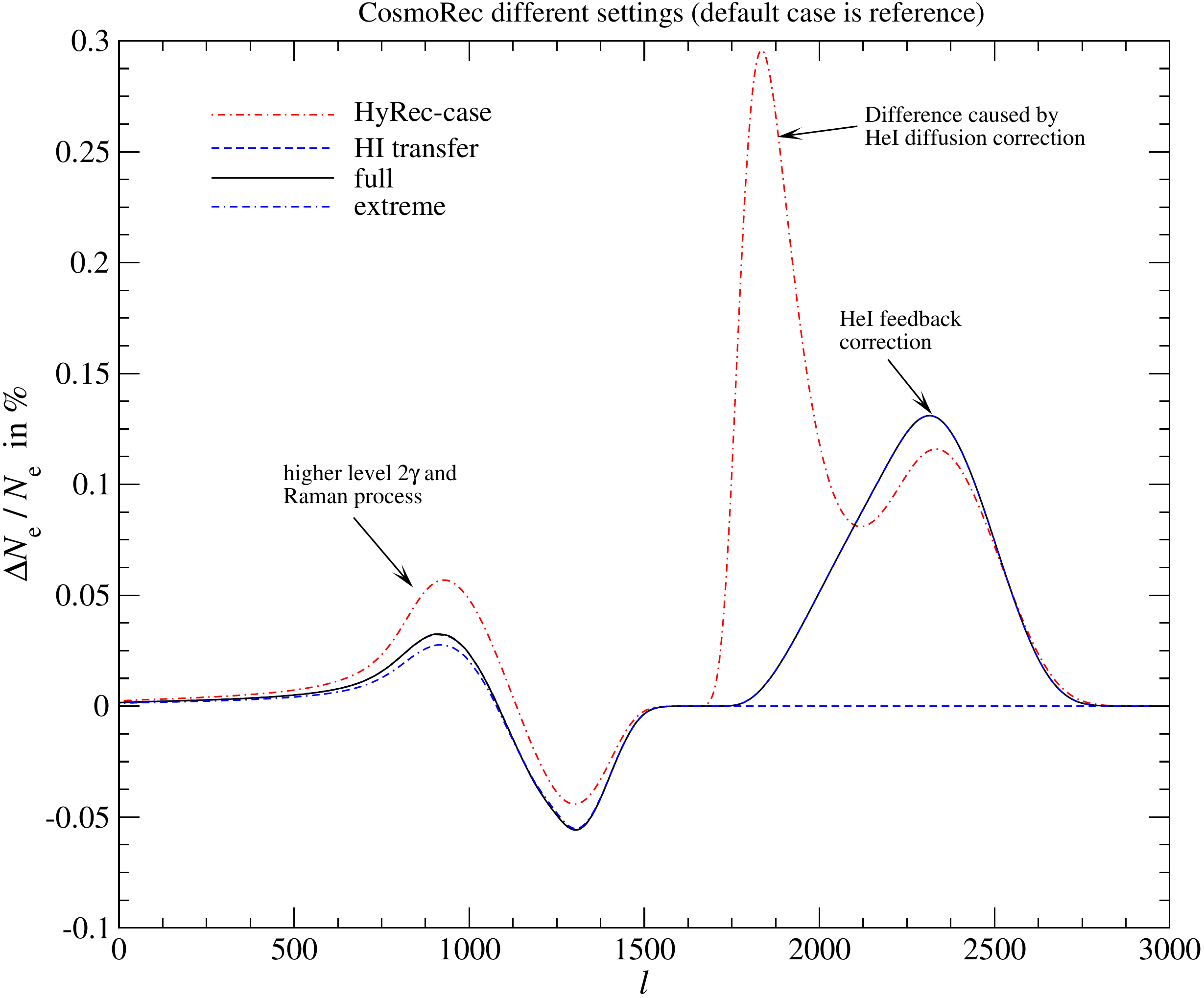}
\caption{Modifications to the cosmological ionization history. -- Left panel: comparison of different recombination models (see Table~\ref{tab:CosmoRec_set}) computed using {\sc CosmoRec} with {\sc Recfast++}. -- Right panel: relative difference between the different recombination models. Here the {\sc CosmoRec} default case is used as reference.
}
\label{fig:Xe}
\end{figure*}

\begin{figure*}
\centering
\includegraphics[width=\columnwidth]{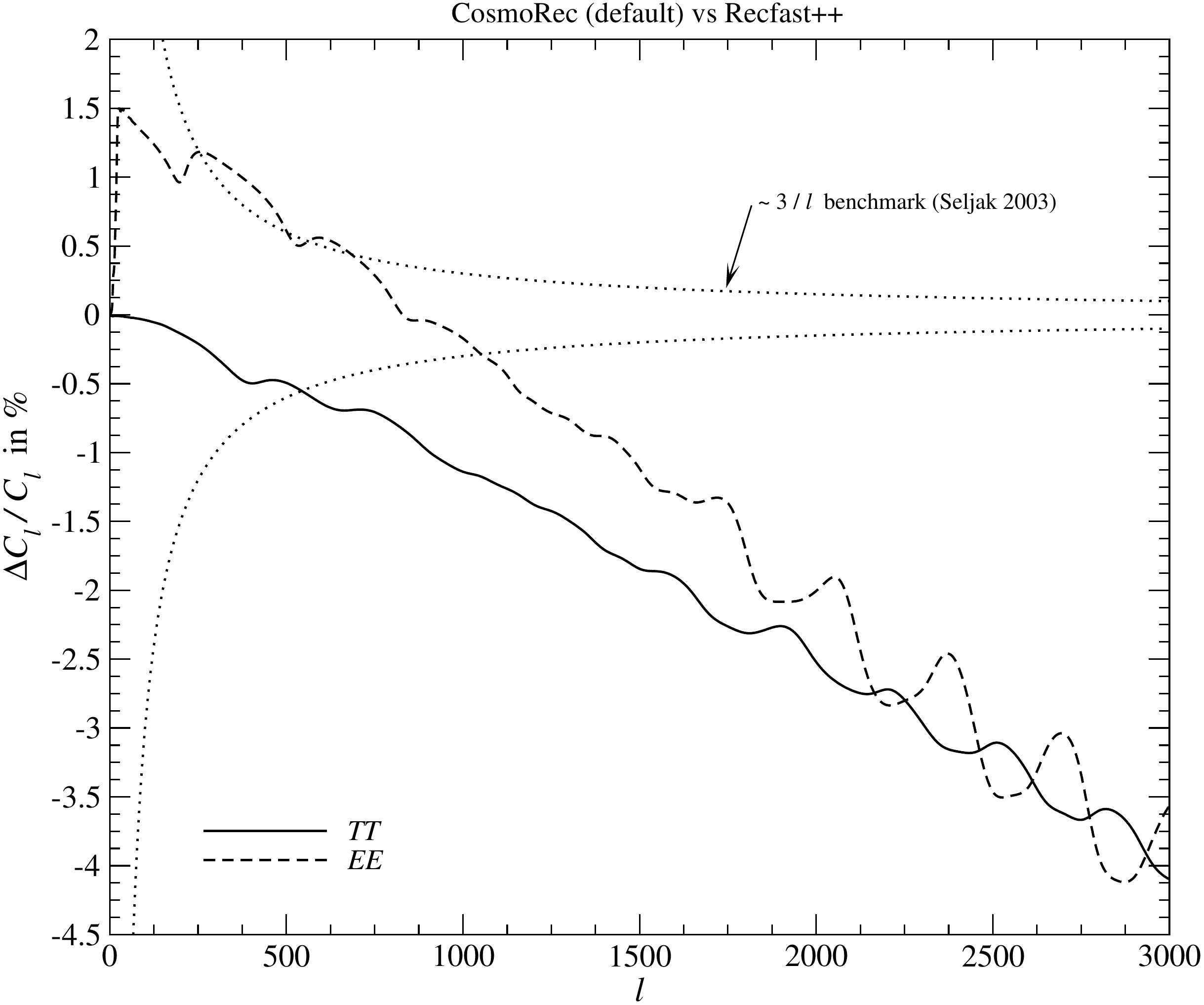}
\includegraphics[width=\columnwidth]{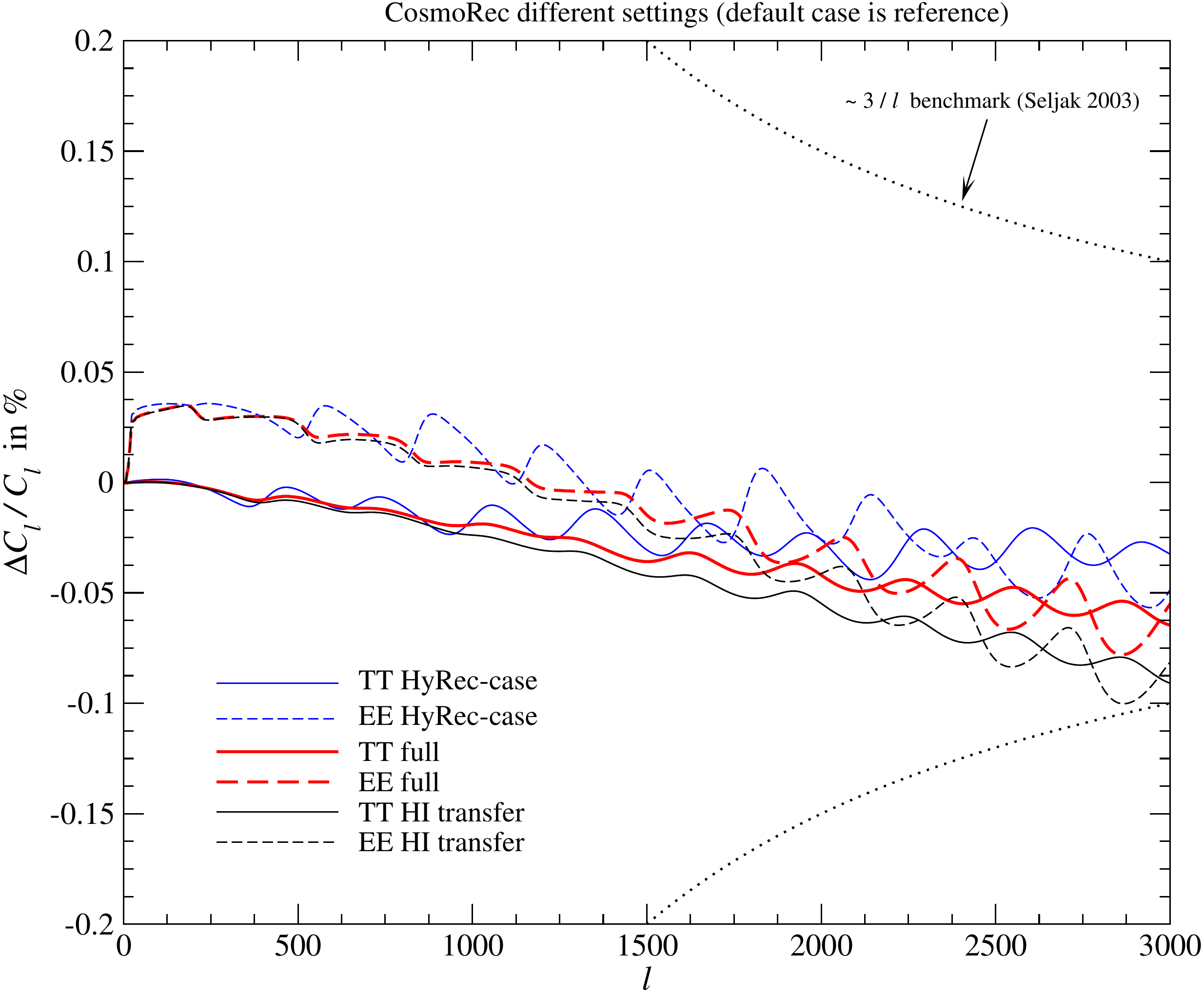}
\caption{Changes in the CMB temperature and polarization power spectra. -- Left panel: correction for the {\sc CosmoRec} default case with respect to the standard {\sc Recfast++} case. -- Right panel: additional corrections for different recombination models. Here the power spectra obtained using the ionization history of the {\sc CosmoRec} default case define the reference. In both panels we show the $3/l$ benchmark \citep{Seljak2003} for comparison (dotted line).
}
\label{fig:DCl}
\end{figure*}

The differences in the free electron fraction \changeJ{with respect to the original
version of {\sc Recfast} \citep{SeagerRecfast1999} reach the level of a
\changeJ{$\Delta N_{\rm e}/N_{\rm e}\sim 1\%-3\%$}}  (see Fig.~\ref{fig:Xe} for more details).
These modifications affect the shape of the Thomson visibility
function \citep{Sunyaev1970}, slightly shifting its position and
changing its width.
This leads to important differences in the theoretical predictions of
the CMB temperature and polarisation power spectra, reaching $\Delta
C_l/C_l \sim -4\%$ at $l\sim 3000$ with respect to \changeJ{the result obtained with the ionisation history given by {\sc Recfast}} (see
Fig.~\ref{fig:DCl}).
Neglecting these corrections leads to biases in the inferred
parameters, which are important at the level of a few standard
deviations for {\Planck} \citep{Jose2010}.

In this paper we address the important question about how the
aforementioned modifications to the ionization history propagate into
different cosmological parameters. We use the recombination code {\sc
  CosmoRec} in connection with {\sc CAMB} \citep{CAMB} and {\sc
  CosmoMC} \citep{COSMOMC} for our computations.

With {\sc CosmoRec} it has at last become possible to perform {\it
  model-by-model} parameter estimations, including {\it all} important
physical corrections to the recombination process in a detailed
recombination calculation.
%
%% HERE
The study by \citet{Jose2010} was partially reliant on the results of
the multi-dimensional interpolation scheme {\sc Rico}
\citep{Fendt2009}, and the final parameter estimations were performed
using the newly introduced {\it correction function} \citep{Jose2010}.
Also, {at that time} the recombination code used for hydrogen \citep{Chluba2007} was
limited to 100 shells only, while it is now possible to use effective
rates \citep{Yacine2010} for up to 500 shells \citep{Chluba2010b},
capturing the correct behaviour in the freeze-out tail of
recombination ($z\lesssim 800$).

Furthermore, we show in more detail how different combinations of
parameters affect the associated biases.
In particular, {in addition to the standard six parameters} we allow for a variation of the helium abundance,
$\YHe$, and the effective number of neutrino species, $\Nnu$, finding
that the main biases are strongly reshuffled (see
Table~\ref{tab:planck} and \ref{tab:cv}).
\changeJ{We also} extend the analysis to forecast possible biases to
combined {\Planck} plus {\sc ACTPol} and {\sc SPTpol} data sets.
\changeJ{Finally, we} address the question of which physical processes are really
important in the recombination correction, and whether an approach
based on a {\it correction function} \citep{Jose2010} is sufficient
for future parameter estimation.

The paper is structured as follows: in Section~\ref{sec:recphys}, {we
describe {\sc CosmoRec} and some of the important corrections} to
the recombination process; in Section~\ref{sec:parameter} we explore
the biases produced by different recombination calculations and their
implications for cosmological data analysis; finally in
Section~\ref{sec:precision} we discuss the {necessary} accuracy of
calculation required for unbiased analysis of the CMB, {providing some details on a {\sc Recfast} correction function approach}.

\section{Recombination physics}
\label{sec:recphys}
\label{sec:std_phys}

In this section we {briefly discuss} the different
recombination corrections included in the computations carried out
with {\sc CosmoRec}\footnote{{\sc CosmoRec} is available at
  \url{http://www.Chluba.de/CosmoRec}}. For a more \changeJ{general} overview on
recombination physics we refer the interested reader to
\citet{Fendt2009}, \citet{Sunyaev2009} and \citet{Jose2010}.

We study the precision of {\sc CosmoRec} using different combinations
of the recombination physics, as well as various settings for the
solver parameters.
The main purpose is to demonstrate that using {\sc CosmoRec} with
default setting is sufficient for precise analysis of \changeJ{{\Planck} data, and even for a combination of {\Planck} plus {\sc ACTpol} and {\sc SPTpol}}. 
We also address this question again in Section~\ref{sec:parameter} using {\sc
  CosmoMC}.
For all results presented in this Section we used the fiducial
cosmological model given in Table~\ref{tab:planck}.
%cosmological parameters $\omegab = 0.0226$, $\omegadm = 0.112$, $T_0=2.726\,$K, $H_0=70.0\, {\rm km\,s^
%{-1}\,Mpc^{-1}}$, $Y_{\rm p}=0.24$, and $N_\nu=3.046$.

%---------------
%---------------

%---------------
%---------------

\subsection{Corrections to the cosmological ionization history and the CMB power spectra}
\label{sec:Xe}
{\sc CosmoRec} allows the incorporation of all known, important,
corrections to the cosmological recombination problem. However, its
runtime varies rather strongly with the level of detail in the
recombination model, and for parameter estimations using {\sc CosmoMC}
it is important to reduce the runtime as far as possible, when using
the full recombination calculation.
It is therefore {useful} to check which corrections need to be
accounted for to obtain unbiased results in the {final} parameter estimation.

In Table~\ref{tab:CosmoRec_set} we define a set of recombination
models to illustrate the possible differences in the ionization
history.
The corresponding corrections to the ionization history as a function
of redshift are presented in Fig.~\ref{fig:Xe}, and Fig.~\ref{fig:DCl}
shows the associated modifications in the CMB power spectra.
For the curves \changeJ{given} in the left panel of Fig.~\ref{fig:Xe} the output
from our C++ version of {\sc Recfast}\footnote{ {\sc Recfast++} is
  part of {\sc CosmoRec} and can also be downloaded separately at
  \url{http://www.Chluba.de/CosmoRec}.
  It reproduces the {\sc Recfast} result \citep{SeagerRecfast1999},
  but avoids any switches in the ODE system \citep[see][for
  details]{Fendt2009}.
  Running {\sc Recfast++} is similar to running {\sc Recfast v1.5}
  \citep{Wong2008} with all hydrogen and helium flags set to zero.  }
was used as reference case, while in the right panel the relative
differences between the alternative models are illustrated, with the
{\sc CosmoRec} `default' case defining the reference.

The left panel of Fig.~\ref{fig:Xe} shows, that the recombination
corrections with respect to the original version of {\sc Recfast}
reach the level of a few percent, both during hydrogen and helium
recombination.
During helium recombination ($z\sim 1700-2200$) the largest correction
($\sim -2\%$) is because of the increase in the photon escape rate of
the $\rm 2^1P-1^1 S$ singlet line {mediated by the absorption of helium photons in the \ion{H}{i}
continuum} \citep{Kholupenko2007, Switzer2007I, Jose2008}.
During hydrogen recombination ($z\sim 800-1500$) details in the
radiative transfer of the Lyman-series are {crucial}, including {\it
  two-photon corrections}, {\it resonance scattering}, and {\it
  Raman-events} \citep[see][\changeJ{and references therein} for more details]{Chluba2010b,
  Yacine2010}.
The modification in the CMB power spectra related to the {\sc
  CosmoRec} `default' model are shown in the left panel of
Fig.~\ref{fig:DCl}.

As the right panel of Fig.~\ref{fig:Xe} indicates, the changes in the
ionization history between the different models of
Table~\ref{tab:CosmoRec_set} are already very small.
For all cases shown the differences in the hydrogen recombination
history are $\Delta N_{\rm e}/ N_{\rm e} \lesssim 0.1\%$.
The largest difference appears when switching off the diffusion
correction to the escape probability of the \ion{He}{i} $\rm 2^1P-1^1
S$ resonance \citep{Jose2008}, resulting in $\Delta N_{\rm e}/ N_{\rm
  e} \sim 0.3\%$ uncertainty at $z\sim 1800$,
a correction that currently is not included by {\sc HyRec}.
During hydrogen recombination, higher level two-photon decays ($n>3$)
do still lead to some $\sim 0.1\%$ uncertainty, which appears to be
dominated by the 4s-1s and 4d-1s process.
However, when computing the corresponding changes in the CMB
temperature and polarization power spectra (right panel of
Fig~\ref{fig:DCl}), it becomes clear that the small difference with
respect to the {\sc CosmoRec} `default' case will not have a major
impact on the cosmological parameter constraints for cosmic variance
limited experiments at $l\lesssim 3000$.
The agreement in the prediction for the CMB power spectra is better
than $0.1\%$ for all considered recombination models.
This error is below the $3/l$ benchmark suggested by
\citet{Seljak2003}, and as we will see below, indeed there is no
significant bias introduced when choosing between the different
recombination models defined in Table~\ref{tab:CosmoRec_set}.

These findings suggests that one can use {\sc CosmoRec} with the
`default' setting to perform accurate cosmological parameter
estimations with {\sc CosmoMC}.
Most importantly, for this setting {\sc CosmoRec} runs in 1.3 seconds
per cosmology, and was already tested in a wide range of cosmological
parameters to assure stability of the recombination code.
In contrast to the most demanding setup (`extreme' case of
Table~\ref{tab:CosmoRec_set}) {\sc CosmoRec} runs about 340 times
faster in this mode.
We also confirmed the precision of {\sc CosmoRec} by comparing
directly with the most detailed computation carried out using a more
elaborate multi-level hydrogen-helium recombination code
\citep{Chluba2010b}.
We found differences no larger than $\Delta N_{\rm e}/ N_{\rm e} \sim
0.01\%$ at all redshifts.

We will address {the question about the precision of the recombination model} more formally in
Section~\ref{sec:precision}, however, we find that for accurate
parameter estimation using future data from {\Planck}, {\sc ACTPol}
and {\sc SPTpol} the {\sc CosmoRec} `default' setting is indeed
sufficient.
\changeJ{However, we would like to point out that a final cross-validation of the {\sc CosmoRec} outputs with independent recombination codes \citep{Switzer2007I, Hirata2008, Grin2009, Yacine2010b} will be very important.
Nevertheless, we do not expect our conclusions to change very much.}

%---------------------------
\begin{table*}
\centering
\caption{Different settings for {\sc CosmoRec}. In all cases the corrections to the \ion{H}{i} 2s-1s two-photon channel were switched on \citep{Chluba2006, Kholu2006}. Furthermore, we used the effective rates for our hydrogen and helium models with $n_{\rm eff}^{\rm HI}=500$ and $n_{\rm eff}^{\rm HeI}=30$ \citep{Chluba2009c, Chluba2010}. {These} were computed \changeJ{with} the method of \citet{Yacine2010}. The \ion{He}{i} $\rm 2^3P-1^1 S$ intercombination line was always switched on \citep{Dubrovich2005} and the effect of \ion{H}{i} continuum absorption on the $\rm 2^1P-1^1 S$ singlet and $\rm 2^3P-1^1 S$ triplet resonance \citep{Kholupenko2007, Switzer2007I} was included using the {\it no redistribution} approximation \citep{Jose2008}. 
In all cases, the detailed history was solved starting at $z=3000$ and ending at $z=50$ with $500$ intermediate points. The solution was completed until $z=0$ using the simple {\sc Recfast} ODE system, unless stated differently.
Depending on the corresponding settings, the feedback of helium photons was treated according to \citet{Chluba2009c}, including $n^1\rm D-1^1 S$ quadrupole lines and $n^3\rm P-1^1 S$ intercombination lines from levels $n>2$. The diffusion correction to the escape probability of the \ion{He}{i} $\rm 2^1P-1^1 S$ resonance was included using the tabulated correction function given in \citet{Jose2008}.
Higher level two-photon decays and Raman-scatterings were incorporated according to \citet{Chluba2010b}, 
{giving results that are in excellent agreement with \citet{Hirata2008}}.
}
\begin{tabular}{lccccc}
\hline
parameter & default  &  {\sc HyRec}-case$^\dagger$ & HI transfer & full  & extreme \\
\hline 
resolved states & 2s-3s, 2p-3p, 3d  & 2s-4s, 2p-4p, 3d-4d 
& 2s-8s, 2p-10p, 3d-8d & 2s-8s, 2p-10p, 3d-8d & 2s-8s, 2p-10p, 3d-8d \\[1mm]
\# resolved states & 5  & 8 & 22 & 22 & 22 \\[1mm]
%
%$n_{\rm eff}^{\rm HI}$ & $500$ & $500$ & 500 & 500 \\[1mm]
%
$n_{2\gamma}$ & $3$ & $4$ & $8$& $8$ & $8$ \\[1mm]
$n_{\rm Raman}$ & $2$ & $3$ & $7$ & $7$ & $7$ \\[1mm]
\hline
$n_{\rm max}^{\rm HeI}$ & $2$ & $2$ & 2 & 5 & 5 \\[1mm]
%
%$n_{\rm eff}^{\rm HeI}$ & $30$ & $30$ & 30 & 30 & 30 \\[1mm]
%
HeI diffusion correction & on & off & on & on & on \\[1mm]
HeI feedback & off & $n_{\rm f}=2$ &  off & $n_{\rm f}=5$ & $n_{\rm f}=5$ \\[1mm]
\hline
$\Delta z$ (PDE-solver) & $20$ & $20$ & 2 & 2 & 0.2\\
average runtime & 1.3 sec & 2.2 sec & 38 sec  & 50 sec & 350 sec \\
\hline
\end{tabular}
\\
$^\dagger$ This case is intended to reproduce the result of {\sc HyRec} \citep{Yacine2010c} using {\sc CosmoRec}.
\label{tab:CosmoRec_set}
\end{table*}
%---------------------------

\subsection{Modifications to the freeze-out tail of recombination}
\label{sec:HI_freeze}
Earlier computations of hydrogen recombination \citep{Chluba2007} were
limited to atomic models with $\sim 100$ shells.
However, it was later shown that, as expected \citep{Chluba2007}, the
freeze-out tail of recombination ($z\sim 800$) is still affected
significantly until about \changeJ{$300-400$} shells are included
\citep{Grin2009, Chluba2010}.
{With} the effective rate method \citep{Yacine2010} it has now
become possible to account for this correction in the computation of
the CMB power spectra {in a very fast way}.

When going from 100 shells to {400} shells, the correction in the
freeze-out tail of hydrogen changes from about $\Delta N_{\rm
  e}/N_{\rm e}\sim 2.8\%$ to $\sim 1.6\%$ \citep[see for example
Fig.~2 in][]{Chluba2010b}.
This \changeJ{is expected to lead to} some small change in the CMB power spectra, since the
total optical depth to the last scattering surface is slightly
{modified}.
However, the difference is very small, and as we show here in detail
(Section~\ref{sec:precision}), for precise parameter estimation with
{\Planck} it would be sufficient to include only 100 shells to the
computation.
Although this was already suggested by \citet{Jose2010}, here we {explicitly} show
this using {\sc CosmoRec}, however, as the effective rate coefficients
can be easily computed even up to 500 shells, this does not lead to
any additional obstacle in the recombination calculation.

The main {implication of this is} that changes at the level of $\sim 1\%$ in
the freeze-out tail of recombination are not constrainable with {\sc
  Planck}.
This further suggests that modifications caused by collisional
processes \citep{Chluba2007, Chluba2010} at $z\lesssim 800$ should not
matter \changeJ{very much}.
It was already shown that collisions lead to a small acceleration of
recombination at $z\lesssim 800$. Since collisional rates are very
uncertain, this could \changeJ{imply modifications} $\gtrsim 0.1\%$
\citep{Chluba2010}. However, since changes $\sim 1\%$ at low redshifts
($z\lesssim 800$) do not seem to affect the CMB power spectra at a
significant level (see Section~\ref{sec:precision}), \changeJ{the above statement}
appears reasonable.
Nevertheless, it will be important to check \changeJ{this} with
refined computations of collisional rate coefficients.

%---------------
\begin{figure}
\centering
\includegraphics[width=\columnwidth]{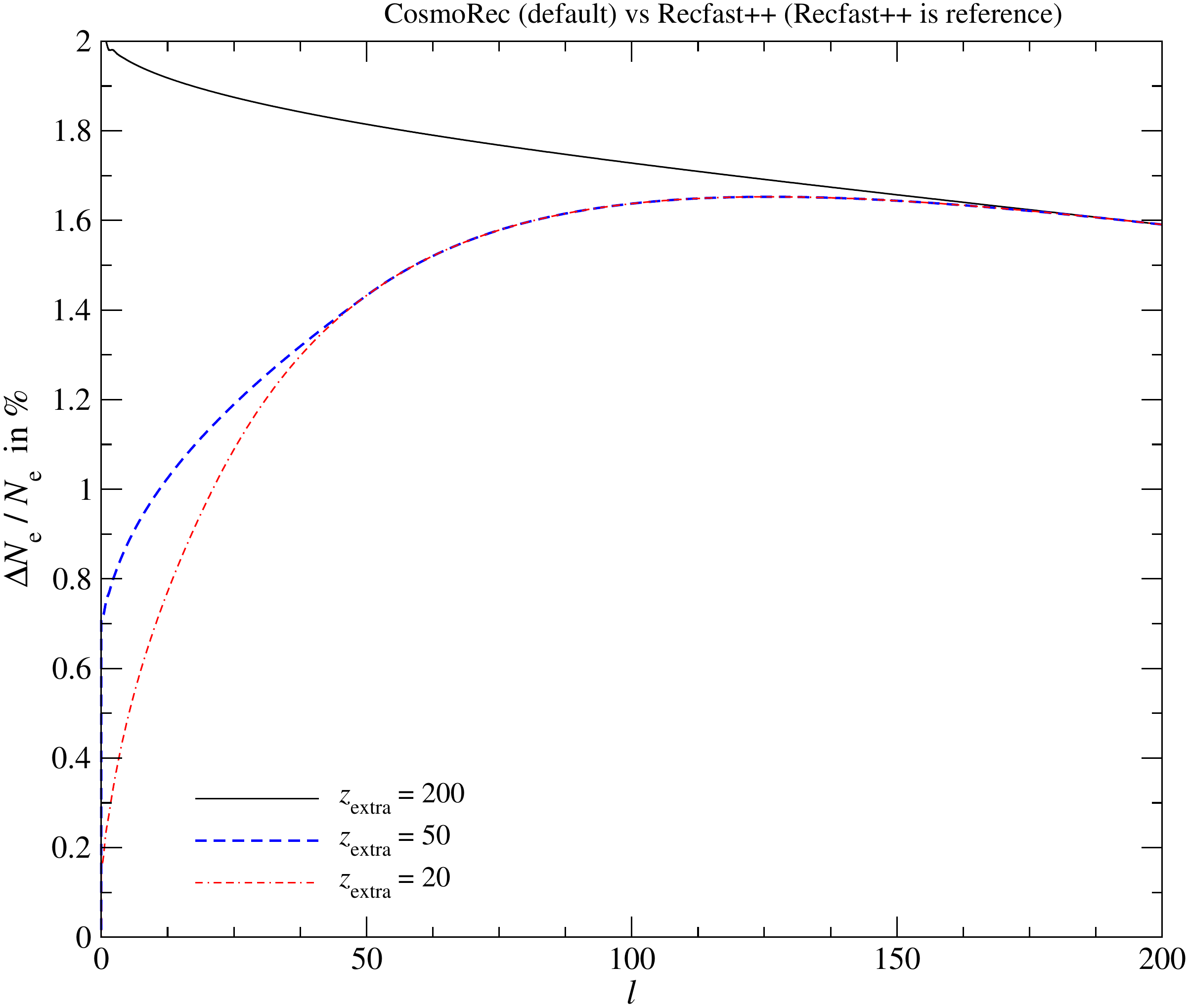}
\caption{Modifications to the cosmological ionization history at low redshifts. We compare the output of {\sc CosmoRec} for default setting with {\sc Recfast++}. The difference in the models is only due to the low-$z$ extrapolation, which is done using the simple {\sc Recfast++} ODE system at $z<z_{\rm extra}$. The models shown here lead to changes in the CMB power spectra with $\Delta C_l/C_l \lesssim 10^{-5}$ at $l\lesssim 3000$.
}
\label{fig:Xe_low}
\end{figure}
%---------------
\subsection{Computation at very low ($z\lesssim 200$) redshifts}
\label{sec:Xe_extra}
In the current version of {\sc CosmoRec} the recombination problem is
solved until some lower redshift, $z_{\rm extra}$. Below this redshift
the simple ODE system of {\sc Recfast++} is used to complete the
solution until $z=0$ \citep{Chluba2010b}.
The derivatives of {\sc CosmoRec} at $z_{\rm extra}$ are used to
re-scale the derivatives of {\sc Recfast++} accordingly.
Clearly this procedure is expected to introduce some small error to
the ionization history, however, as Fig.~\ref{fig:Xe_low} demonstrates
the differences are small, when varying the value of $z_{\rm extra}$.
In particular, we found that the associated uncertainty in the CMB
power spectra for all shown cases is $\Delta C_l/C_l \lesssim 10^{-5}$
at $l\lesssim 3000$.
For {\sc CosmoRec} we shall therefore use $z_{\rm extra}=50$ for all
computations, without introducing any significant change to the CMB
power spectra.

This result also suggests that any modifications introduced by details
in the primordial chemistry \citep{Stancil1996, Stancil1998,
  Schleicher2008} should not have a major impact on the predictions
for the CMB power spectra, as they are expected to be similarly small.
{Furthermore, at $z\lesssim 10-20$ the process of cosmological
  reionisation \citep{Barkana2001} after the formation of first stars
  in the Universe \citep[e.g. see][]{Tegmark1997, Abel1997, Abel2000,
    Yoshida2007} is expected to occur. This leads to a much larger
  ambiguity \citep[e.g. see][for model-independent
  estimation]{Mortonson2008} in the cosmological ionization history
  than inherent to detailed recombination calculations. \changeJ{In addition},
  possible changes caused by energy injection from decaying or
  annihilating relic particles \citep[e.g. see][]{Chen2004,
    Padmanabhan2005, Zhang2006, Zhang2007, Huetsi2009} and/or cosmic
  rays \citep[e.g.][]{Jasche2007}, could affect the ionization history
  at these low redshifts.
Overall the uncertainty at low redshifts will be dominated by other processes than recombination physics, \changeJ{and detailed analysis of their relevance is beyond the scope of this paper}.}

%\subsection{Effect of dark matter annihilation}
%\label{sec:DM_effect}

\section{Parameter estimation using {\sc CosmoRec}}
\label{sec:parameter}

The difference between \changeJ{ionization history obtained with the
  original version of {\sc Recfast} \citep{SeagerRecfast1999} and
  \textsc{CosmoRec}\footnote{This work was performed using
    \textsc{CosmoRec} version 1.3b.} leads} to differences in the
statistics of the CMB that become increasingly important at small
scales (see Section~\ref{sec:recphys}). Though these differences are
tiny, in any experiment the smallest cosmic variance limited scales
are given a huge weight in the likelihood function, and may lead to
significant biasing when inferring cosmological parameters. Our task
in this {Section} is to assess the importance of the deviations in
recombination calculations on the analysis of future cosmological
experiments --- our primary approach to this will be to analyse
simulated data for various experiments using different recombination
calculations and deduce how well the input model is reproduced. We
will compare three different recombination calculations
\textsc{CosmoRec}, \textsc{Recfast++} and \textsc{Recfast++} using {a
  correction function \citep{Jose2010}}, which was obtained using the
{`full'} model given in Table~\ref{tab:CosmoRec_set} for a cosmology
close to the fiducial model (given in Table~\ref{tab:planck}).
{We will henceforth call \textsc{Recfast++} with correction
  function \textsc{Recfast++cf}.}

For this work we rely on \textsc{CosmoMC} \citep{COSMOMC}, a
sophisticated Markov-Chain Monte-Carlo (MCMC) package that has become
the standard code for the analysis of cosmological data. At its core
is the Boltzmann code \textsc{CAMB} \citep{CAMB} which calculates
cosmological statistics given the proposed set of cosmological
parameters, and this uses the \textsc{Recfast v1.5} system for
calculating the baryon ionisation fraction and baryon temperature. At
the simplest level, incorporating \textsc{CosmoRec} into the
\textsc{CosmoMC} stack is a simple matter of modifying \textsc{CAMB}
to replace the \textsc{Recfast} calculation. We have made these
modifications publicly available\footnote{The modified code and
  instructions can be found at \url{http://camb.info/jrs/cosmorec/}.}.

In this work we have performed the cosmological analysis using several
different models, all based around the standard six parameter
$\Lambda$CDM model. For parameters we use the dimensionless baryon and
dark matter densities, $\Omega_b h^2$ and $\Omega_c h^2$; the optical
depth to last scattering $\tau$; the spectral index of the primordial
power spectrum $n_s$ and the logarithm of its amplitude $\log{(10^{10}
  A_s)}$ (both defined at a pivot scale of $k_0 = 0.05
\mathrm{Mpc}^{-1}$); finally we will use the Hubble parameter $H_0$
(instead of $\theta$).  In choosing this parameterisation we have
enforced a flat universe. We are neglecting any secondary
contributions to the CMB --- including those from the
Sunyaev--Zel'dovich effect, gravitational lensing, and inhomogeneous
reionisation --- as they should be small at the scales we are
considering (that is $l < 2000$ in all but one case).

The additional models we will consider are all extensions to
$\Lambda$CDM: allowing the primordial Helium fraction $\YHe$ to vary;
having a variable number of neutrino species $N_\nu$ (all of which we
assume to be massless); and giving the freedom to fit a more complex
primordial power spectrum by adding the running of the spectral index
$n_\text{run}$ as a parameter.
\changeJ{CMB based constraints on $\YHe$ and $\Nnu$ were recently obtained by \citet{Dunkley2010} using combined {\sc Act} and {\sc Wmap} data, demonstrating the power of small scale with full sky experiments. In the future such constraints will tighten very much, and as we show here the recombination corrections with become very important in this case.}

For \changeJ{our} simulated data we use an exact realisation of the CMB angular
powerspectrum calculated for our input model using \changeJ{the}
\texttt{all\_l\_exact} functionality of \textsc{CosmoMC}\footnote{See
  \url{http://cosmocoffee.info/viewtopic.php?t=231}.}. The noise
properties of the simulated data are detailed in subsequent
sections. We use no data (simulated or otherwise) in addition to this
CMB data. Our input powerspectrum is calculated using the `full'
accuracy setting of \textsc{CosmoRec} (see
Table~\ref{tab:CosmoRec_set}) and with all \textsc{CAMB}'s accuracy
level flags set to 4. However when performing the analysis we use the
`default' \textsc{CosmoRec} accuracy; we give a detailed discussion of
the implications of this {choice} in
Section~\ref{sec:precision}.

For analysing cosmic-variance limited data up to a {multipole}
of $l = 2000$ the precision of \textsc{CosmoMC} must be increased to
at least \texttt{accuracy\_level = 2}. This {ensures} that it
is able to reproduce the simulated data at the precision demanded by
the cosmic-variance limit. If the default accuracy {of {\sc
    CosmoMC}} is used, a log-likelihood of {$\mathcal{O}(1)$}
is obtained for the input model (which should have a log-likelihood of
zero).
In each case discussed we have used at least four chains for the MCMC
sampling. We have ensured tested their convergence using the
Gelman-Rubin \citep{GelmanRubin} statistic, in all cases terminating
when $R-1 < 0.005$.

\subsection{WMAP}
\label{sec:parameter_WMAP}

We have performed an analysis of the WMAP 7-year data release
\citep{Komatsu2010} on its own for the models listed above, using the
three different recombination calculations. As expected from the work
of \cite{Jose2010}, there were no observable distinctions between the
three, and we confirm that the recombination corrections are not
important for WMAP.

\subsection{Parameter estimation for \textsc{Planck}}
\label{sec:parameter_Planck}

\textsc{Planck} is expected to be cosmic variance limited up until $l
\sim 1500$ for temperature observations, (though E-mode polarisation
is affected by instrumental noise at all scales). This increased
precision means it is much more sensitive to the recombination
corrections {than {\sc WMAP}} \citep{Jose2010}.

We have both repeated the analysis of \citep{Jose2010}, and
{also} extended it by considering models with $\YHe$ and
$N_\nu$ included as free parameters. In the simulated data we simply
presume that the {final \textsc{Planck} map} is full sky with
the same overall noise properties as the $143 \: \mathrm{GHz}$ band
(assuming the other frequency bands have been used to clean
foregrounds). This is the same approximation as \cite{Jose2010}. This
means we use a beam scale of $\theta_\text{beam} = 7.1'$, and noise of
$\sigma_T^2 \Omega_\text{beam} = 1.53 \times 10^{-4} \:
\mu\mathrm{K}^2$ and $\sigma_P^2 \Omega_\text{beam} = 5.59 \times
10^{-4} \: \mu\mathrm{K}^2$.

\begin{table*}
\centering
\caption{The differing recombination calculations could significantly
  affect the analysis of Planck data. The table below illustrates the
  biases using each of three different recombination codes, in both
  their absolute deviation and the number of sigmas. Deviations over
  $1\sigma$ are highlighted. For the bias in sigmas we use multiples 
  of the standard deviation of each distribution, a definition that is 
  robust to their non-gaussianity. Though a more desirable alternative would 
  be to turn the confidence limit for each into an effective number 
  of sigmas, this would be highly unreliable as the tails are very 
  sparsely sampled at the large deviations observed.}

\input{plancktable.tex}

\label{tab:planck}
\end{table*}

The results of our simulations are summarised in
Table~\ref{tab:planck}. For each parameter in the models listed we
give its fiducial value, the value constrained by \textsc{CosmoRec}
with it standard error, {followed by} the absolute and relative bias for
each of the three recombination calculations.
\changeJ{We now discuss specific cases in more detail.}

\subsubsection{$\Lambda$CDM and $\Lambda$CDM with running}
In the standard $\Lambda$CDM
case we reproduce the results of \cite{Jose2010} finding significant
biases in the parameters $\Omega_b h^2$ and $n_s$ when recovered by
\textsc{Recfast++} ($\sim 2 \sigma$ and $3 \sigma$ respectively). From
Fig.~\ref{fig:DCl} we can see that the main effect of the
recombination corrections on the $C_l$'s is to lower the damping tail
further, in light of this it is understandable that the bias obtained
shifts $n_s$ in order to lower the small scale multipoles. 
{We note that the biases reported here include the {\it total} correction 
to the ionization history with respect to the original version of 
{\sc Recfast} \citep{SeagerRecfast1999}.}
{In \cite{Jose2010}, part of the computations were performed using 
{\sc Recfast v1.4.2}, which allows to account for part of the corrections to 
helium recombination \citep{Wong2008}, using fudge factors.
Importantly, in comparison to the original version of {\sc Recfast} we find significant shifts} 
for the $\Lambda$CDM + running model in the parameters $\Omega_b h^2$, $n_s$
and $n_\text{run}$ (shifts of at least $1 \sigma$, $3 \sigma$ and $1
\sigma$ respectively), that were not reported in \cite{Jose2010}.

\subsubsection{$\Lambda$CDM with varying $\Yp$ and $\Nnu$}
In the $\Lambda$CDM + Helium model we have allowed variation in the
primordial Helium fraction {$\Yp$} rather than simply relying
on the results from Big Bang Nucleosynthesis {\citep[see][for
  recent constraints and discussion]{Cyburt2003, Steigman2009}}.
Running {parameter estimations with the \textsc{Recfast++}
  reveals} a large discrepancy in $\YHe$ compared to the fiducial
model, recovering a value over $2\sigma$ higher. As increasing the
Helium fraction removes a larger proportion of electrons prior to
hydrogen recombination, this change can partially mimic the lower
electron fraction found in a full calculation (see Fig.~\ref{fig:Xe}).
{In fact, virtually all other biases disappear when freely
  varying $\Yp$. With $\sim 0.5\sigma$ only $H_0$ is significantly
  biased in addition for this case.  }

The CMB is sensitive to the effective number of neutrino species
$N_\nu$ primarily because they affect the background radiation density
of the Universe. {Any significant} deviation from the
theoretical value of $3.046$ \citep{Mangano2005} could indicate a
further generation of neutrinos, or potentially some other source of
background radiation.  As illustrated in
{Table~\ref{tab:planck}}, to infer this value correctly
requires an accurate calculation of the recombination history. To do
otherwise can lead to a bias of up to $1.6 \sigma$ in the recovered
value (as shown by using \textsc{Recfast++}), as well as a similar
shift in $\Omega_b h^2$, and a slightly smaller change in $H_0$. The
high dimensionality of the parameter space, and the numerous shifts
makes it difficult to attribute them to any definite physical effect.
We also allow both $\YHe$ and $N_\nu$ as free parameters, in an eight
parameter model. This results in a smaller set of shifts, with a
higher value of $\YHe$ {acquiring} the most significant shift (at
{$1.3\sigma$}).

\subsubsection{Parameter estimation using \textsc{Recfast++cf}}
{As the results in Table~\ref{tab:planck} indicate, at} the
precision of this \textsc{Planck} simulation there is no significant
{difference between the recovered result} for either of the
\textsc{CosmoRec} or \textsc{Recfast++cf} cases for any
{cosmological} model.
This confirms that a correction function approach in principle is
sufficient for the analysis of {\Planck} data, as already pointed
out by \citet{Jose2010}.  However, given that one run of {\sc
  CosmoRec} with `default' setting takes $\sim 1.3$ seconds or about
$\sim 15\%$ of the total CPU time of {\sc CAMB} per
model\footnote{This is with a timing of 7.5 seconds per \textsc{CAMB}
  call (at \texttt{accuracy\_level = 2}) on a single processor. Using
  \textsc{OpenMP}, \textsc{CAMB} can be parallelised to reduce \changeJ{its}
  overall time, \changeJ{however, {\sc CosmoRec} does not benefit from parallelisation 
  at the moment}.}, it has now become possible to explicitly run the
full recombination code, without large penalty.
% %

\begin{table*}
\centering
\caption{Deviations between the different recombination calculations 
  for an experiment cosmic-variance limited until $l = 2000$ (except in the final case listed). The 
  relative deviations are significantly larger than those in a \textsc{Planck} 
  like experiment (Table~\ref{tab:planck}). Any deviations over 
  $2\sigma$ have been highlighted.}

\input{cvtable.tex}

\label{tab:cv}
\end{table*}

\begin{figure*}
\centering
\includegraphics[width=0.96\textwidth]{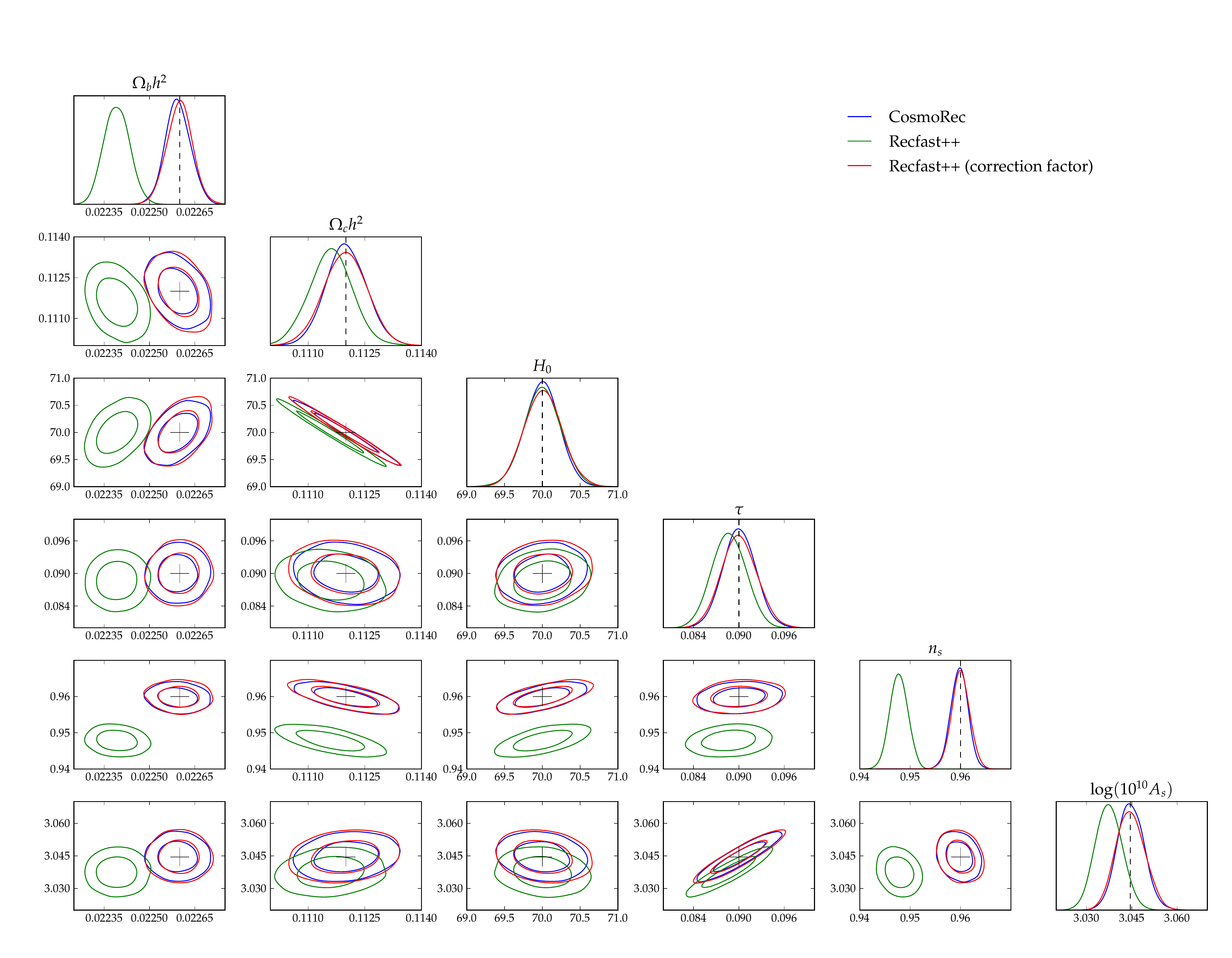}
\caption{Recovered constraints from a simulated experiment, cosmic
  variance limited in both temperature and polarisation until
  $l_\text{max} = 2000$. This plot shows the constraints obtained for
  the six parameter model, with the dashed lines and crosses marking
  the input values in the one- and two-dimensional plots. The contours
  mark the 68\% and 95\% confidence limits of each distribution. The
  three cases illustrated show that using both \textsc{CosmoRec} and
  corrected \textsc{Recfast++} produce nearly identical probability
  distributions which faithfully recover the input model. In contrast,
  the standard \textsc{Recfast++} model significantly biases the
  parameters (most notably $\omegab$ and $n_s$).}
\label{fig:cv_std6}
\end{figure*}

\begin{figure*}
\centering
\includegraphics[width=0.96\textwidth]{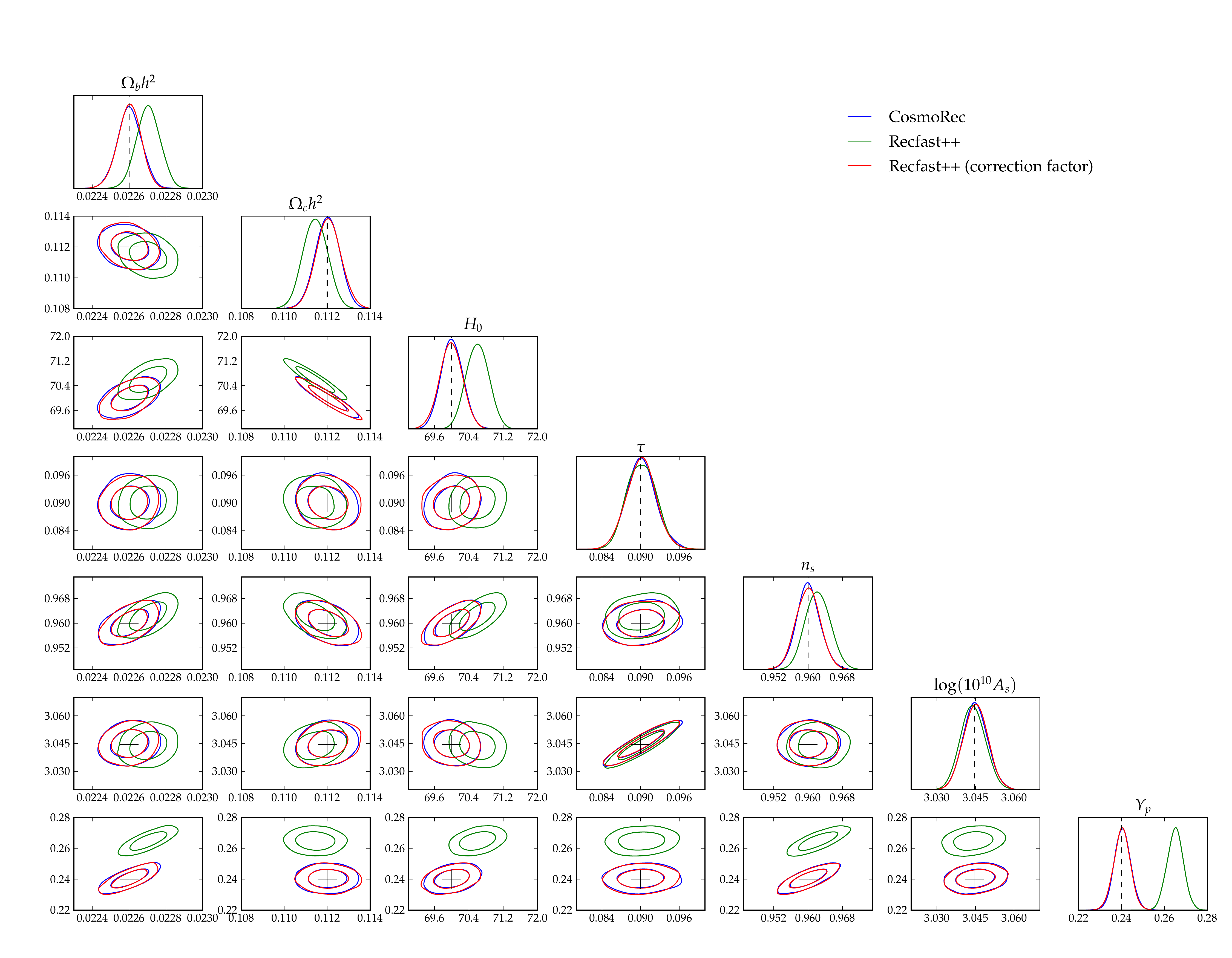}
\caption{Constraints with our cosmic-variance limited experiment for
  $l < 2000$, for the seven-parameter model of $\Lambda$CDM plus an
  unknown primordial Helium content, $\YHe$. The extra degree of freedom
  shifts the bias of the standard \textsc{Recfast++} calculation into
  a dramatically higher value of $\YHe$.}
\label{fig:cv_std6+he}
\end{figure*}

\subsection{Cosmic variance limited parameter estimation}
\label{sec:parameter_CV}

Future experiments such as {\textsc{SPTpol} and
  \textsc{ACTPol} \changeJ{will be much more sensitive than
  \textsc{Planck} at small scales}.}
{This means that the combined \textsc{Planck} plus
  \textsc{SPTpol} and \textsc{ACTPol} data sets} will be sample
variance dominated to much smaller scales than currently accessible
\footnote{Up to $l \sim 2000$ for $E$-mode polarisation in the case of
\textsc{Planck} \changeJ{plus the} \textsc{ACTPol wide} survey.
\textsc{ACTPol deep} may go as far as $l \sim 4000$. See
\citet{ACTPol}}. Rather than exploring the consequences of the
recombination corrections for any particular experiment, or
combination of
experiments, we will use a simulated, full-sky, experiment cosmic variance limited
until \changeJ{some} maximum multipole used in our analysis, usually $l =
2000$. Apart from this we perform a similar analysis to the
{one of the} previous Section. Our results are summarised in
Table~\ref{tab:cv}.

Overall the {obtained set of biases for each combination of
  parameters} are similar to the \textsc{Planck} cases shown
previously. However, the relative shifts are much greater in every
case. For instance in the plain $\Lambda$CDM model, the bias increases
to over 5 and $7\sigma$ for $\Omega_b h^2$ and $n_s$ respectively
(also see Fig.~\ref{fig:cv_std6}). In this case the significant driver
of this increase is the overall tightening of the constraints from the
added data. However, in many cases there is also an absolute shifting
of the biased parameters. This happens as the weight afforded to the
largest multipoles eliminates the previously favourable regions,
shifting them further away. This is most obvious in the shift of $H_0$
in the $\Lambda$CDM + Helium model (see Fig.~\ref{fig:cv_std6+he} as
well as Table~\ref{tab:cv}).

In Table~\ref{tab:cv} we also include the results from the
$\Lambda$CDM model using simulated data from an experiment cosmic
variance limited up to $l = 3000$. In this case the uncertainty on
$\Omega_b h^2$ reduces by around a half, which increases the
{associated} bias to over $10\sigma$. {For $n_s$,}
however, the distribution width remains comparable, but an increase in
the absolute shift, moves the bias to over $12\sigma$.

{Here it is important to mention that at $l\sim 3000$ the
  additional ambiguities caused by SZ clusters, are expected to mainly
  impact the $TT$ power spectra, while leaving the $EE$ power spectra
  less contaminated.
  This holds the potential that foreground subtraction and the removal
  of secondary anisotropies up to $l\sim 3000$ could become feasible
  in the future, in particular for the $E$-mode power spectrum.
  As our results clearly show, in such cases the recombination
  corrections will be extremely important.  }

% \begin{figure*}
% \centering
% \includegraphics[width=0.96\textwidth]{./pdf/cv_std6+nu.pdf}
% \caption{Parameter constraints for the seven-parameter model of
%   $\Lambda$CDM plus a variable number of massless neutrinos
%   $N_\nu$. This is still with our default input model, and an
%   experiment cosmic-variance limited for $l < 2000$. The biases in the
%   \textsc{Recfast++} case are significantly different to those of
%   Fig.~\ref{fig:cv_std6} favouring a significantly larger Hubble
%   constant and dark matter fraction as well as an increased number of
%   neutrinos.}
% \label{fig:cv_std6+nu}
% \end{figure*}

% \begin{figure*}
% \centering
% \includegraphics[width=0.96\textwidth]{./pdf/cv_std6+run.pdf}
% \caption{The seven-parameter model of $\Lambda$CDM with a running
%   spectral index --- again with our simulated cosmic-variance limited
%   experiment. The standard \textsc{Recfast++} case exhibits similar
%   biases to the $\Lambda$CDM case (see Fig.~\ref{fig:cv_std6}), but
%   additionally shows a significantly negative value of $n_\text{run}$
%   (the input dataset had no running).}
% \label{fig:cv_std6+run}
% \end{figure*}

\section{Effect of precision in the recombination physics}
\label{sec:precision}

{In Section \ref{sec:parameter} we have used {\sc CosmoRec}
  with the `default' setting to perform model-by-model computations of
  the recombination process. However, it is clear that this leads to
  some residual errors with respect to the most precise recombination
  calculation (model `extreme' in Table~\ref{tab:CosmoRec_set}).
  In terms of the CMB power spectra this 'approximation' leads to very
  small uncertainties, as discussed already in
  Section~\ref{sec:recphys} (see Fig.~\ref{fig:DCl} for details).
  Here we ask at what level these differences actually matter for
  future CMB parameter estimation, using explicit runs of {\sc
    CosmoRec}.  }

\subsection{Changes to the freeze-out tail of recombination}
\begin{figure}
\centering
\includegraphics[width=\columnwidth]{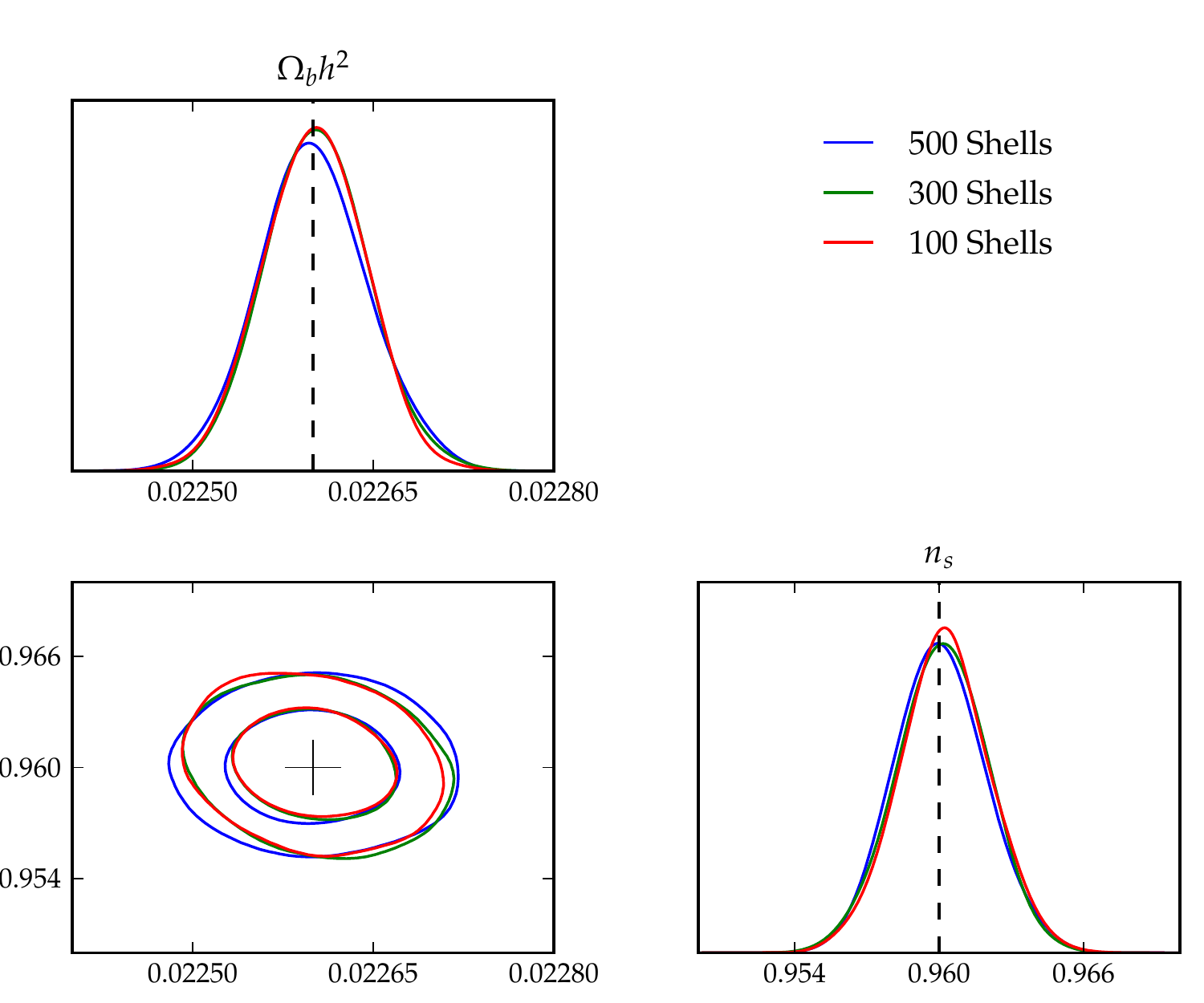}
\caption{The projection of the $\Omega_b h^2$--$n_s$ plane from the
  6-parameter $\Lambda$CDM model, analysed using a maximum of $100$,
  $300$ and $500$ shells in \textsc{CosmoRec}. As can be seen for an
  experiment cosmic-variance limited at $l = 2000$ or less, using a
  maximum of $100$ shells is sufficient.}
\label{fig:shells_2pars}
\end{figure}

In Section~\ref{sec:HI_freeze} we discussed the effect of the number
of shells calculated on the low redshift ionisation fraction. We
expect the relative insensitivity of the CMB power spectrum, means
that the number of shells included is not of great importance.
Using \textsc{CosmoRec} this can be explicitly verified 
{by varying the number of hydrogen levels included into 
the computation of the effective rate coefficients}.
In Fig.~\ref{fig:shells_2pars} we can see that even in the cosmic
variance \changeJ{limited} case there is no significant bias introduced when reducing
the number of shells down to 100. For \textsc{Planck} we expect $100$
shells to be sufficient, confirming {the analysis of}
\cite{Jose2010}.

\subsection{{\sc CosmoRec} `default' versus `full' setting}
As remarked upon in the previous section, our simulated data has been
generated with the `full' setting of \textsc{CosmoRec}, which takes
around 50 {seconds} to evaluate the recombination history.
However, this is too slow for use during an MCMC analysis, {as
  it would make one execution of {\sc CAMB} even with highest accuracy
  setting about seven times slower.}
By using the `default' setting we are introducing differences in the
power spectrum of around $0.05 \%$ (at $l \sim 2000$;
{cf. Fig.~\ref{fig:DCl}}), though small, this induces slight
changes to likelihood surface, shifting the preferred parameters. This
means that even the \textsc{CosmoRec} case exhibits slight biases when
looking at cosmic variance limited data (see
Table~\ref{tab:cv}). However, these are limited to around $0.15
\sigma$. This deviation can be reduced by running \textsc{CosmoRec} at
higher accuracy. Using the `HyRec' setting \changeJ{(which does not include the \ion{He}{i} diffusion
  correction, but takes 4s/4d-1s two-photon decay and 3s/3d-1s Raman events into account; see Table~\ref{tab:CosmoRec_set} for more details)} the maximum bias is reduced
down to $0.09 \sigma$, and if {the \ion{He}{i} diffusion
  corrections is also included} this \changeJ{decreases} to $0.06 \sigma$.
{In both cases the runtime of {\sc CosmoRec} is increased only
  by \changeJ{about $1$ second}, a relatively small penalty.
  We conclude \changeJ{that for {\Planck}} parameter analysis the {\sc
    CosmoRec} `default' setting is sufficient, however, if needed
  additional corrections can be included without compromising the
  computation time very much.  
  \changeJ{Furthermore, our estimation shows that the \ion{He}{i} diffusion
  correction leads to a small additional modification, which can be safely neglected.}}

\subsection{Iterative use of {\sc Recfast++cf}}
\label{sec:iterate}
One alternative to using an explicit \textsc{CosmoRec} calculation
\changeJ{(or any other more detailed recombination code)} is to
{resort to} an incremental approach, {based on {\sc
    Recfast++} with correction function}.
In the absence of any {prior} knowledge about the appropriate
cosmology, this would commence with an initial analysis of the data
using \textsc{Recfast++}.
Then the best fit parameters are used to generate a correction
function for that cosmology {using the most precise
  recombination calculation \citep[e.g.][]{Chluba2010b, Yacine2010c}},
which should be correct within a nearby region in parameter space. The
data is then re-analysed with \textsc{Recfast++} using this correction
function. This process can be iterated until the results are
convergent.

\begin{figure}
\centering
\includegraphics[width=\columnwidth]{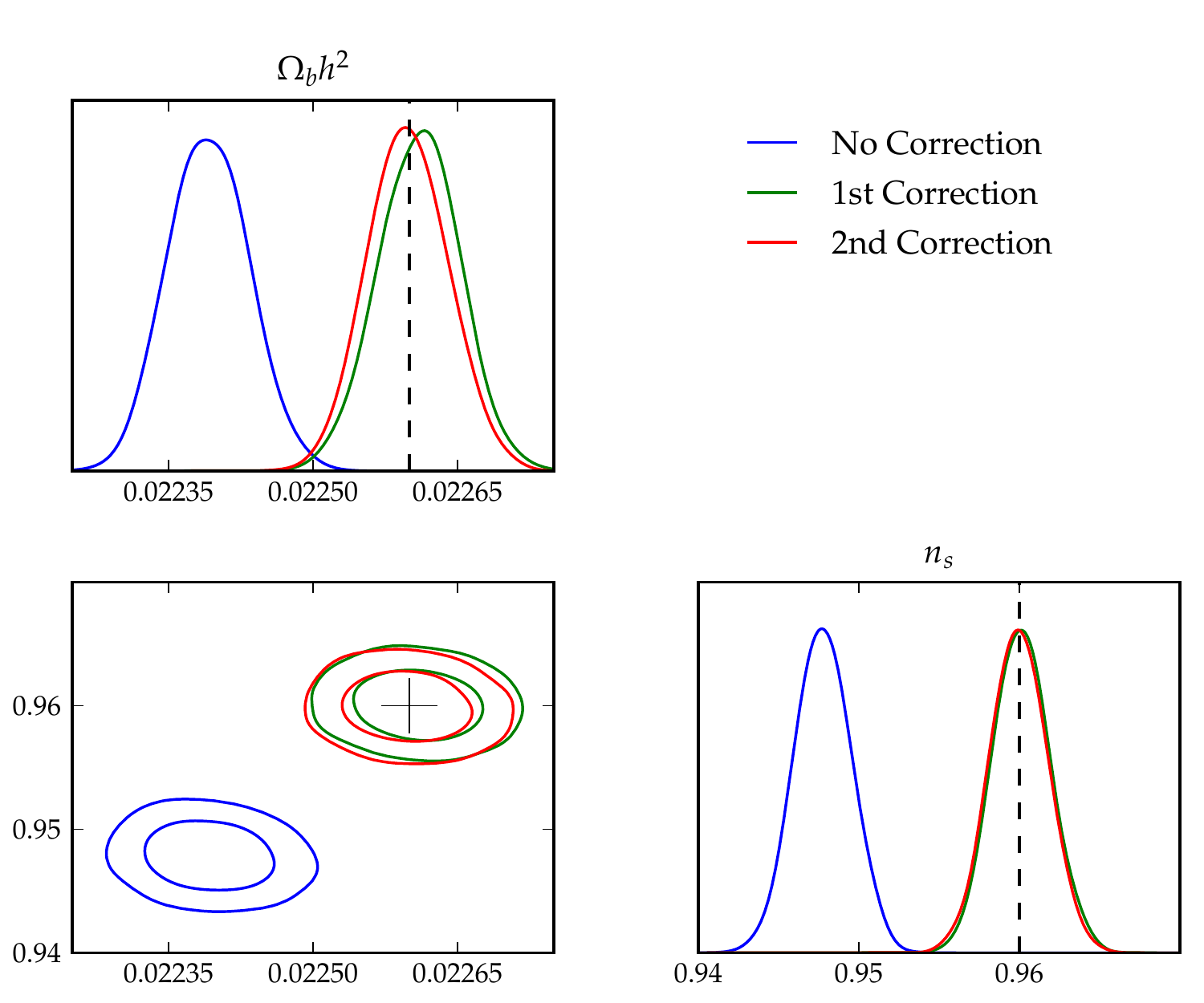}
\caption{Successive iterations of the \textsc{Recfast++} correction
  function in the $\Omega_b h^2$--$n_s$ plane (taken from a full
  six-parameter $\Lambda$CDM analysis). The `No Correction' plot,
  calculated using \textsc{Recfast++}, {exhibits significant
    biases}, however, from this we can calculate the first correction
  function. This correction function can be used in
  \textsc{Recfast++}, to produce {improved} constraints. This
  iteration is much less biased, with the largest change exhibited a
  $0.26 \sigma$ shift in $\Omega_b h^2$. A further iteration with a
  recalculated correction function, removes the last remaining
  biases.}
\label{fig:cf_2pars}
\end{figure}

We have shown the results of this process in Fig.~\ref{fig:cf_2pars},
where we illustrate the $\Omega_b h^2$--$n_s$ plane (projected from
the full $\Lambda$CDM parameter space), using the cosmic-variance,
$l_\text{max} = 2000$ simulated data. The initial results from
\textsc{Recfast++} are highly biased, as discussed in the previous
section. However the first iteration significantly corrects this, with
the largest residual being the $0.26\sigma$ shift in $\Omega_b h^2$, \changeJ{indicating that 'corrections-to-corrections' are very small already}. A
further iteration reduces this shift down to only $0.03 \sigma$,
suggesting that in this cases two iterations are sufficient. This
could likely be reduced down to only one iteration if our cosmological
knowledge is sufficient to calculate a suitable initial correction
function, eliminating the \textsc{Recfast++} only step.

{Furthermore, at the initial step one could use {\sc Recfast
    v1.5} that mimics some of the recombination corrections using {\it
    fudge functions} that were calibrated on the results of earlier
  precise recombination calculations \citep{Jose2010}.
  However, since the fudge function approach \citep{Wong2008} in
  essence is equivalent to a correction function approach
  \citep{Jose2010}, one does not expect any additional improvement in
  this case.
  Once we are unwilling to run improved recombination codes (i.e. {\sc
    CosmoRec} or {\sc HyRec}), a correction function approach will
  lead to a very similar approximation at re-calibrated fudge
  functions to {\sc Recfast}.  }

\section{Conclusions}
\label{sec:conc}
%---------------------------
In this work we have demonstrated in detail \changeJ{how previously neglected physical
processes during the recombination epoch affect the analysis of precise CMB data from \textsc{Planck} and beyond.}
\changeJ{We have successfully incorporated {\sc CosmoRec} into the widely used parameter estimation code {\sc CosmoMC} and the Boltzmann code {\sc Camb}.}
\changeJ{With this we were able to} show the parameter biases introduced in a comprehensive set of models
when \emph{neglecting} these corrections. 
We confirm the significant biases in the standard six-parameter $\Lambda$CDM model previously \changeJ{reported by} \cite{Jose2010} for \changeJ{case of} \Planck. In addition to considering a wider range of cosmological models, we also extend this \changeJ{investigation} to a future dataset (comparable to \textsc{Planck} plus \textsc{ACTPol}/\textsc{SPTpol}) using full model-by-model computations of the ionization history with {\sc CosmoRec}  (see Tables~\ref{tab:planck} and \ref{tab:cv} for details).
There are several biases of particular note, which have not been mentioned elsewhere:

\begin{itemize}

\item Allowing a running of the spectral index yields a bias \changeJ{of} over $1\sigma$  in $n_{\rm run}$ for \textsc{Planck}. For an experiment  that is noise-free up to $l = 2000$ this increases to over $3\sigma$.

\item If the primordial Helium fraction is to be inferred using CMB data alone
it may be biased by up $7\sigma$. If uncorrected this would \changeJ{erroneously} indicate a
 significant tension with \changeJ{predictions from} Big-Bang Nucleosynthesis.

\item \changeJ{Adding $\Nnu$ as a free parameter would bias its estimated value high by up to $1.6\sigma$ for \Planck, however, when also varying the helium fraction this decreases to $0.3\sigma$ only.
}

\end{itemize}

Our analysis indicates that for the current precision of combined {\sc WMAP} plus {\sc ACT} data, the bias in $\YHe$ corresponds to about $0.5\sigma$. This is small compared to the $\sim 2\sigma$ tension with the BBN value reported by \cite{Dunkley2010}.
However, in the near future the error in the measurement should decrease and the corresponding bias in $\YHe$ caused by a neglect of detailed recombination physics should exceed the level of $1\sigma$, even prior to the final release of \Planck  data.

Although small scale CMB data is able to internally break many of the
degeneracies of current cosmological analyses, it will still be
desirable to supplement future CMB data with complementary probes such
as Type Ia supernovae data \citep{Amanullah2010} or BAO measurements \citep{BigBOSS}. We have not included
any additional data in this work, and whilst it may serve to reduce
the overall biases obtained (if using a \textsc{Recfast} analysis),
the tension between datasets is likely to artificially tighten
parameter constraints.

In order to focus on the importance of the recombination corrections
we have necessarily neglected many instrumental and physical
systematics:
\begin{itemize}
\item Both the SZ effect and gravitational lensing will significantly
 add to the small scale CMB. We have neglected these in our analysis
 assuming that for $l < 2000$ they will be negligible or sufficiently
 well understood.

\item Our work uses a simple reionisation model which is assumed to be
 known (other than a single parameter giving the optical depth). A
 more complicated homogeneous reionisation history should only affect
 the large scale multipoles, whereas any realistic inhomogeneity will not
 significantly effect multipoles \changeJ{$l \sim 2000$} \citep[see e.g.,][]{Zahn2005}.

\item We have assumed a simple, perfectly known, noise model in this
 work. In reality beam, gain and other uncertainties will have to be
 considered \citep[see e.g.][]{Colombo2009}.

\end{itemize}

The `default' accuracy setting for \textsc{CosmoRec} is designed to be
a fast enough to be used for MCMC analysis of cosmological data, and
this work shows that it is sufficiently accurate for the unbiased
analysis of \textsc{Planck} data. We also discuss an iterative
approach using \textsc{Recfast++} with a correction function. Though
this approach is accurate (see Sect.~\ref{sec:iterate}), it is simpler and quicker overall to directly 
use the full \textsc{CosmoRec} analysis. 
For the analysis of future \changeJ{\textsc{Planck} plus \textsc{ACTPol}/\textsc{SPTpol}} data, 
we show that \textsc{CosmoRec} with `default' setting is unbiased at the
$0.15\sigma$ level, which can be significantly reduced by running at a
higher accuracy setting (leading to a small runtime increase).

It is important to note that a final cross-validation of the {\sc
 CosmoRec} outputs with independent recombination codes
\citep{Switzer2007I, Hirata2008, Grin2009, Yacine2010b} will be very
important, for confirming their veracity.
Nevertheless, initial comparisons \changeJ{mean that we do not expect any
significant changes (at least for the experiments simulated
in this work).}

\changeJ{We would also like to mention that in this work we have demonstrated the importance of the {\it total} corrections to the ionization history with respect to the original {\sc Recfast} code \citep{SeagerRecfast1999}. This correction was obtained in a common effort by several independent groups, and part of these corrections are now included by the latest version of {\sc Recfast} using calibrated correction functions, based on the more detailed recombination computations.
  As we demonstrated here, such an approach is similar to an iterative
  scheme using a precomputed correction function for the original {\sc
    Recfast} computation (see Sect.~\ref{sec:iterate}). However, we
  also argued that for runtimes achieved with recent recombination
  codes, such an approximate approach can be avoided.  }

\section*{Acknowledgments}
The authors would like to thank Yacine Ali-Ha{\"i}moud, Dick Bond, Dan
Grin, Amir Hajian, Chris Hirata, Antony Lewis, Jose-Alberto
Rubi{\~n}o-Mart{\'{\i}}n, Mike Nolta, Hiranya Peiris, Douglas Scott,
Rashid Sunyaev and Eric Switzer for useful and stimulating discussions
of the problem.
JC is also very grateful for additional financial support from
the Beatrice~D.~Tremaine fellowship 2010.  
Furthermore, we acknowledge the use of the GPC supercomputer at the
SciNet HPC Consortium. SciNet is funded by: the Canada Foundation for
Innovation under the auspices of Compute Canada; the Government of
Ontario; Ontario Research Fund - Research Excellence; and the
University of Toronto.

\bibliographystyle{mn2e}
\bibliography{Lit}

%---------------
\end{document}

%% file: Befehle.tex
\newcommand{\beq}{\begin{equation}}   %

\newcommand{\eeq}{\end{equation}}   %

\newcommand{\beqa}{\begin{eqnarray}}   %

\newcommand{\eeqa}{\end{eqnarray}}   %

\newcommand{\beal}{\begin{align}}

\newcommand{\enal}{\end{align}}

\newcommand{\bspl}{\begin{split}}

\newcommand{\espl}{\end{split}}

\newcommand{\bsub}{\begin{subequations}}

\newcommand{\esub}{\end{subequations}}

\newcommand{\bmulti}{\begin{multline}}   %

\newcommand{\beqm}{\begin{mathletters}}   %

\newcommand{\eeqm}{\end{mathletters}}   %

\newcommand{\Yp}{Y_{\rm p}}

%% file: plancktable.tex
\begin{tabular}{ccc|cc|cc|cc}
\toprule
Parameters & Fiducial & Recovered & \multicolumn{2}{c}{\textsc{CosmoRec}} &\multicolumn{2}{c}{\textsc{Recfast++}} & \multicolumn{2}{c}{\textsc{Recfast++} w/ correction} \\
& & & Absolute & Sigmas & Absolute & Sigmas & Absolute & Sigmas \\
\midrule
\multicolumn{9}{l}{$\Lambda$CDM} \\[2mm]
      $\Omega_b h^2$ & $      0.0226$ & $          \ensuremath{0.02260} \pm           \ensuremath{0.00014}$  &           \ensuremath{-0.0000} &             \ensuremath{-0.01}  &  \ensuremath{\mathbf{-0.0003}} &    \ensuremath{\mathbf{-2.10}}  &            \ensuremath{0.0000} &              \ensuremath{0.01} \\
      $\Omega_c h^2$ & $       0.112$ & $           \ensuremath{0.1121} \pm            \ensuremath{0.0012}$  &            \ensuremath{0.0001} &              \ensuremath{0.05}  &            \ensuremath{0.0006} &              \ensuremath{0.48}  &            \ensuremath{0.0001} &              \ensuremath{0.04} \\
               $H_0$ & $          70$ & $            \ensuremath{69.98} \pm              \ensuremath{0.60}$  &           \ensuremath{-0.0200} &             \ensuremath{-0.03}  &           \ensuremath{-0.4800} &             \ensuremath{-0.78}  &           \ensuremath{-0.0200} &             \ensuremath{-0.03} \\
              $\tau$ & $        0.09$ & $           \ensuremath{0.0905} \pm            \ensuremath{0.0045}$  &            \ensuremath{0.0005} &              \ensuremath{0.10}  &           \ensuremath{-0.0017} &             \ensuremath{-0.40}  &            \ensuremath{0.0003} &              \ensuremath{0.07} \\
               $n_s$ & $        0.96$ & $           \ensuremath{0.9598} \pm            \ensuremath{0.0036}$  &           \ensuremath{-0.0002} &             \ensuremath{-0.06}  &  \ensuremath{\mathbf{-0.0120}} &    \ensuremath{\mathbf{-3.35}}  &           \ensuremath{-0.0001} &             \ensuremath{-0.03} \\
 $\log(10^{10} A_s)$ & $      3.0445$ & $           \ensuremath{3.0456} \pm            \ensuremath{0.0089}$  &            \ensuremath{0.0011} &              \ensuremath{0.12}  &  \ensuremath{\mathbf{-0.0091}} &    \ensuremath{\mathbf{-1.05}}  &            \ensuremath{0.0010} &              \ensuremath{0.11} \\
\midrule
\multicolumn{9}{l}{$\Lambda$CDM + He} \\[2mm]
      $\Omega_b h^2$ & $      0.0226$ & $          \ensuremath{0.02260} \pm           \ensuremath{0.00020}$  &            \ensuremath{0.0000} &              \ensuremath{0.02}  &   \ensuremath{2\times 10^{-5}} &              \ensuremath{0.10}  &  \ensuremath{-1\times 10^{-5}} &             \ensuremath{-0.04} \\
      $\Omega_c h^2$ & $       0.112$ & $           \ensuremath{0.1120} \pm            \ensuremath{0.0013}$  &           \ensuremath{-0.0000} &             \ensuremath{-0.03}  &           \ensuremath{-0.0002} &             \ensuremath{-0.17}  &            \ensuremath{0.0001} &              \ensuremath{0.05} \\
               $H_0$ & $          70$ & $            \ensuremath{70.03} \pm              \ensuremath{0.72}$  &            \ensuremath{0.0300} &              \ensuremath{0.05}  &            \ensuremath{0.3800} &              \ensuremath{0.51}  &           \ensuremath{-0.0300} &             \ensuremath{-0.04} \\
              $\tau$ & $        0.09$ & $           \ensuremath{0.0905} \pm            \ensuremath{0.0045}$  &            \ensuremath{0.0005} &              \ensuremath{0.10}  &            \ensuremath{0.0001} &              \ensuremath{0.01}  &            \ensuremath{0.0003} &              \ensuremath{0.06} \\
               $n_s$ & $        0.96$ & $           \ensuremath{0.9601} \pm            \ensuremath{0.0069}$  &            \ensuremath{0.0001} &              \ensuremath{0.02}  &            \ensuremath{0.0004} &              \ensuremath{0.05}  &           \ensuremath{-0.0003} &             \ensuremath{-0.04} \\
 $\log(10^{10} A_s)$ & $      3.0445$ & $           \ensuremath{3.0454} \pm            \ensuremath{0.0095}$  &            \ensuremath{0.0009} &              \ensuremath{0.10}  &           \ensuremath{-0.0017} &             \ensuremath{-0.18}  &            \ensuremath{0.0008} &              \ensuremath{0.08} \\
               $Y_p$ & $        0.24$ & $            \ensuremath{0.240} \pm             \ensuremath{0.011}$  &            \ensuremath{0.0000} &              \ensuremath{0.01}  &   \ensuremath{\mathbf{0.0230}} &     \ensuremath{\mathbf{2.12}}  &           \ensuremath{-0.0010} &             \ensuremath{-0.05} \\
\midrule
\multicolumn{9}{l}{$\Lambda$CDM + Neutrinos} \\[2mm]
      $\Omega_b h^2$ & $      0.0226$ & $          \ensuremath{0.02262} \pm           \ensuremath{0.00021}$  &   \ensuremath{2\times 10^{-5}} &              \ensuremath{0.08}  &           \ensuremath{-0.0000} &             \ensuremath{-0.02}  &   \ensuremath{1\times 10^{-5}} &              \ensuremath{0.04} \\
      $\Omega_c h^2$ & $       0.112$ & $           \ensuremath{0.1123} \pm            \ensuremath{0.0026}$  &            \ensuremath{0.0003} &              \ensuremath{0.11}  &   \ensuremath{\mathbf{0.0044}} &     \ensuremath{\mathbf{1.63}}  &            \ensuremath{0.0001} &              \ensuremath{0.05} \\
               $H_0$ & $          70$ & $             \ensuremath{70.2} \pm               \ensuremath{1.5}$  &            \ensuremath{0.2000} &              \ensuremath{0.12}  &   \ensuremath{\mathbf{2.0000}} &     \ensuremath{\mathbf{1.21}}  &            \ensuremath{0.1000} &              \ensuremath{0.05} \\
              $\tau$ & $        0.09$ & $           \ensuremath{0.0907} \pm            \ensuremath{0.0047}$  &            \ensuremath{0.0007} &              \ensuremath{0.14}  &           \ensuremath{-0.0001} &             \ensuremath{-0.03}  &            \ensuremath{0.0005} &              \ensuremath{0.10} \\
               $n_s$ & $        0.96$ & $           \ensuremath{0.9609} \pm            \ensuremath{0.0081}$  &            \ensuremath{0.0009} &              \ensuremath{0.11}  &            \ensuremath{0.0006} &              \ensuremath{0.07}  &            \ensuremath{0.0004} &              \ensuremath{0.05} \\
 $\log(10^{10} A_s)$ & $      3.0445$ & $            \ensuremath{3.046} \pm             \ensuremath{0.012}$  &            \ensuremath{0.0020} &              \ensuremath{0.16}  &            \ensuremath{0.0040} &              \ensuremath{0.33}  &            \ensuremath{0.0010} &              \ensuremath{0.10} \\
             $N_\nu$ & $       3.046$ & $             \ensuremath{3.07} \pm              \ensuremath{0.19}$  &            \ensuremath{0.0200} &              \ensuremath{0.13}  &   \ensuremath{\mathbf{0.3300}} &     \ensuremath{\mathbf{1.62}}  &            \ensuremath{0.0100} &              \ensuremath{0.05} \\
\midrule
\multicolumn{9}{l}{$\Lambda$CDM + Neutrinos + He} \\[2mm]
      $\Omega_b h^2$ & $      0.0226$ & $          \ensuremath{0.02261} \pm           \ensuremath{0.00023}$  &   \ensuremath{1\times 10^{-5}} &              \ensuremath{0.05}  &   \ensuremath{5\times 10^{-5}} &              \ensuremath{0.24}  &   \ensuremath{1\times 10^{-5}} &              \ensuremath{0.06} \\
      $\Omega_c h^2$ & $       0.112$ & $           \ensuremath{0.1126} \pm            \ensuremath{0.0038}$  &            \ensuremath{0.0006} &              \ensuremath{0.17}  &            \ensuremath{0.0009} &              \ensuremath{0.24}  &            \ensuremath{0.0005} &              \ensuremath{0.13} \\
               $H_0$ & $          70$ & $             \ensuremath{70.3} \pm               \ensuremath{1.8}$  &            \ensuremath{0.3000} &              \ensuremath{0.18}  &            \ensuremath{0.9000} &              \ensuremath{0.52}  &            \ensuremath{0.3000} &              \ensuremath{0.14} \\
              $\tau$ & $        0.09$ & $           \ensuremath{0.0905} \pm            \ensuremath{0.0046}$  &            \ensuremath{0.0005} &              \ensuremath{0.11}  &            \ensuremath{0.0001} &              \ensuremath{0.03}  &            \ensuremath{0.0004} &              \ensuremath{0.10} \\
               $n_s$ & $        0.96$ & $           \ensuremath{0.9608} \pm            \ensuremath{0.0083}$  &            \ensuremath{0.0008} &              \ensuremath{0.10}  &            \ensuremath{0.0019} &              \ensuremath{0.24}  &            \ensuremath{0.0007} &              \ensuremath{0.08} \\
 $\log(10^{10} A_s)$ & $      3.0445$ & $            \ensuremath{3.046} \pm             \ensuremath{0.012}$  &            \ensuremath{0.0020} &              \ensuremath{0.16}  &            \ensuremath{0.0000} &              \ensuremath{0.03}  &            \ensuremath{0.0020} &              \ensuremath{0.13} \\
             $N_\nu$ & $       3.046$ & $             \ensuremath{3.10} \pm              \ensuremath{0.27}$  &            \ensuremath{0.0500} &              \ensuremath{0.19}  &            \ensuremath{0.0900} &              \ensuremath{0.33}  &            \ensuremath{0.0400} &              \ensuremath{0.15} \\
               $Y_p$ & $        0.24$ & $            \ensuremath{0.238} \pm             \ensuremath{0.015}$  &           \ensuremath{-0.0020} &             \ensuremath{-0.13}  &   \ensuremath{\mathbf{0.0190}} &     \ensuremath{\mathbf{1.26}}  &           \ensuremath{-0.0020} &             \ensuremath{-0.11} \\
\midrule
\multicolumn{9}{l}{$\Lambda$CDM + Running} \\[2mm]
      $\Omega_b h^2$ & $      0.0226$ & $          \ensuremath{0.02260} \pm           \ensuremath{0.00015}$  &            \ensuremath{0.0000} &              \ensuremath{0.03}  &  \ensuremath{\mathbf{-0.0002}} &    \ensuremath{\mathbf{-1.40}}  &            \ensuremath{0.0000} &              \ensuremath{0.01} \\
      $\Omega_c h^2$ & $       0.112$ & $           \ensuremath{0.1119} \pm            \ensuremath{0.0013}$  &           \ensuremath{-0.0001} &             \ensuremath{-0.05}  &            \ensuremath{0.0006} &              \ensuremath{0.45}  &            \ensuremath{0.0000} &              \ensuremath{0.00} \\
               $H_0$ & $          70$ & $            \ensuremath{70.04} \pm              \ensuremath{0.62}$  &            \ensuremath{0.0400} &              \ensuremath{0.06}  &           \ensuremath{-0.4100} &             \ensuremath{-0.65}  &            \ensuremath{0.0100} &              \ensuremath{0.01} \\
              $\tau$ & $        0.09$ & $           \ensuremath{0.0905} \pm            \ensuremath{0.0046}$  &            \ensuremath{0.0005} &              \ensuremath{0.11}  &           \ensuremath{-0.0000} &             \ensuremath{-0.01}  &            \ensuremath{0.0004} &              \ensuremath{0.08} \\
               $n_s$ & $        0.96$ & $           \ensuremath{0.9601} \pm            \ensuremath{0.0036}$  &            \ensuremath{0.0001} &              \ensuremath{0.02}  &  \ensuremath{\mathbf{-0.0121}} &    \ensuremath{\mathbf{-3.42}}  &            \ensuremath{0.0000} &              \ensuremath{0.01} \\
 $\log(10^{10} A_s)$ & $      3.0445$ & $           \ensuremath{3.0454} \pm            \ensuremath{0.0096}$  &            \ensuremath{0.0009} &              \ensuremath{0.09}  &           \ensuremath{-0.0043} &             \ensuremath{-0.44}  &            \ensuremath{0.0009} &              \ensuremath{0.10} \\
    $n_\mathrm{run}$ & $           0$ & $          \ensuremath{-0.0000} \pm            \ensuremath{0.0049}$  &           \ensuremath{-0.0000} &             \ensuremath{-0.01}  &  \ensuremath{\mathbf{-0.0058}} &    \ensuremath{\mathbf{-1.17}}  &            \ensuremath{0.0001} &              \ensuremath{0.01} \\
\bottomrule
\end{tabular}

%% file: cvtable.tex
\begin{tabular}{ccccccccc}
\toprule
Parameters & Fiducial & Recovered & \multicolumn{2}{c}{\textsc{CosmoRec}} &\multicolumn{2}{c}{\textsc{Recfast++}} & \multicolumn{2}{c}{\textsc{Recfast++} w/ correction} \\
& & & Absolute & Sigmas & Absolute & Sigmas & Absolute & Sigmas \\
\midrule
\multicolumn{9}{l}{$\Lambda$CDM} \\[2mm]
      $\Omega_b h^2$ & $      0.0226$ & $         \ensuremath{0.022594} \pm          \ensuremath{0.000039}$  &  \ensuremath{-6\times 10^{-6}} &             \ensuremath{-0.15}  &  \ensuremath{\mathbf{-0.0002}} &    \ensuremath{\mathbf{-5.35}}  &  \ensuremath{-1\times 10^{-6}} &             \ensuremath{-0.01} \\
      $\Omega_c h^2$ & $       0.112$ & $          \ensuremath{0.11201} \pm           \ensuremath{0.00051}$  &   \ensuremath{1\times 10^{-5}} &              \ensuremath{0.01}  &           \ensuremath{-0.0004} &             \ensuremath{-0.77}  &  \ensuremath{-2\times 10^{-5}} &             \ensuremath{-0.04} \\
               $H_0$ & $          70$ & $            \ensuremath{69.99} \pm              \ensuremath{0.22}$  &           \ensuremath{-0.0100} &             \ensuremath{-0.03}  &            \ensuremath{0.0000} &              \ensuremath{0.00}  &            \ensuremath{0.0100} &              \ensuremath{0.05} \\
              $\tau$ & $        0.09$ & $           \ensuremath{0.0901} \pm            \ensuremath{0.0020}$  &            \ensuremath{0.0001} &              \ensuremath{0.03}  &           \ensuremath{-0.0013} &             \ensuremath{-0.64}  &            \ensuremath{0.0000} &              \ensuremath{0.02} \\
               $n_s$ & $        0.96$ & $           \ensuremath{0.9598} \pm            \ensuremath{0.0016}$  &           \ensuremath{-0.0002} &             \ensuremath{-0.14}  &  \ensuremath{\mathbf{-0.0122}} &    \ensuremath{\mathbf{-7.48}}  &            \ensuremath{0.0000} &              \ensuremath{0.01} \\
 $\log(10^{10} A_s)$ & $      3.0445$ & $           \ensuremath{3.0446} \pm            \ensuremath{0.0042}$  &            \ensuremath{0.0001} &              \ensuremath{0.04}  &           \ensuremath{-0.0071} &             \ensuremath{-1.72}  &            \ensuremath{0.0000} &              \ensuremath{0.01} \\
\midrule
\multicolumn{9}{l}{$\Lambda$CDM + He} \\[2mm]
      $\Omega_b h^2$ & $      0.0226$ & $         \ensuremath{0.022602} \pm          \ensuremath{0.000059}$  &   \ensuremath{2\times 10^{-6}} &              \ensuremath{0.03}  &            \ensuremath{0.0001} &              \ensuremath{1.78}  &   \ensuremath{1\times 10^{-6}} &              \ensuremath{0.02} \\
      $\Omega_c h^2$ & $       0.112$ & $          \ensuremath{0.11201} \pm           \ensuremath{0.00051}$  &   \ensuremath{1\times 10^{-5}} &              \ensuremath{0.02}  &           \ensuremath{-0.0006} &             \ensuremath{-1.06}  &   \ensuremath{6\times 10^{-5}} &              \ensuremath{0.11} \\
               $H_0$ & $          70$ & $            \ensuremath{70.01} \pm              \ensuremath{0.23}$  &            \ensuremath{0.0100} &              \ensuremath{0.03}  &   \ensuremath{\mathbf{0.6000}} &     \ensuremath{\mathbf{2.50}}  &           \ensuremath{-0.0100} &             \ensuremath{-0.06} \\
              $\tau$ & $        0.09$ & $           \ensuremath{0.0901} \pm            \ensuremath{0.0021}$  &            \ensuremath{0.0001} &              \ensuremath{0.04}  &            \ensuremath{0.0002} &              \ensuremath{0.08}  &            \ensuremath{0.0001} &              \ensuremath{0.04} \\
               $n_s$ & $        0.96$ & $           \ensuremath{0.9601} \pm            \ensuremath{0.0025}$  &            \ensuremath{0.0001} &              \ensuremath{0.03}  &            \ensuremath{0.0023} &              \ensuremath{0.87}  &           \ensuremath{-0.0000} &             \ensuremath{-0.00} \\
 $\log(10^{10} A_s)$ & $      3.0445$ & $           \ensuremath{3.0449} \pm            \ensuremath{0.0043}$  &            \ensuremath{0.0004} &              \ensuremath{0.10}  &           \ensuremath{-0.0005} &             \ensuremath{-0.10}  &            \ensuremath{0.0006} &              \ensuremath{0.13} \\
               $Y_p$ & $        0.24$ & $           \ensuremath{0.2405} \pm            \ensuremath{0.0035}$  &            \ensuremath{0.0005} &              \ensuremath{0.13}  &   \ensuremath{\mathbf{0.0250}} &     \ensuremath{\mathbf{7.24}}  &            \ensuremath{0.0002} &              \ensuremath{0.07} \\
\midrule
\multicolumn{9}{l}{$\Lambda$CDM + Neutrinos} \\[2mm]
      $\Omega_b h^2$ & $      0.0226$ & $         \ensuremath{0.022597} \pm          \ensuremath{0.000058}$  &  \ensuremath{-3\times 10^{-6}} &             \ensuremath{-0.05}  &   \ensuremath{5\times 10^{-5}} &              \ensuremath{0.93}  &  \ensuremath{-3\times 10^{-6}} &             \ensuremath{-0.06} \\
      $\Omega_c h^2$ & $       0.112$ & $          \ensuremath{0.11203} \pm           \ensuremath{0.00099}$  &   \ensuremath{3\times 10^{-5}} &              \ensuremath{0.03}  &   \ensuremath{\mathbf{0.0045}} &     \ensuremath{\mathbf{4.77}}  &           \ensuremath{-0.0000} &             \ensuremath{-0.00} \\
               $H_0$ & $          70$ & $            \ensuremath{69.99} \pm              \ensuremath{0.45}$  &           \ensuremath{-0.0100} &             \ensuremath{-0.02}  &   \ensuremath{\mathbf{2.2800}} &     \ensuremath{\mathbf{5.40}}  &           \ensuremath{-0.0600} &             \ensuremath{-0.14} \\
              $\tau$ & $        0.09$ & $           \ensuremath{0.0902} \pm            \ensuremath{0.0021}$  &            \ensuremath{0.0002} &              \ensuremath{0.08}  &           \ensuremath{-0.0002} &             \ensuremath{-0.11}  &            \ensuremath{0.0000} &              \ensuremath{0.02} \\
               $n_s$ & $        0.96$ & $           \ensuremath{0.9600} \pm            \ensuremath{0.0027}$  &            \ensuremath{0.0000} &              \ensuremath{0.01}  &            \ensuremath{0.0007} &              \ensuremath{0.26}  &           \ensuremath{-0.0002} &             \ensuremath{-0.08} \\
 $\log(10^{10} A_s)$ & $      3.0445$ & $           \ensuremath{3.0449} \pm            \ensuremath{0.0049}$  &            \ensuremath{0.0004} &              \ensuremath{0.08}  &            \ensuremath{0.0066} &              \ensuremath{1.37}  &            \ensuremath{0.0002} &              \ensuremath{0.03} \\
             $N_\nu$ & $       3.046$ & $            \ensuremath{3.046} \pm             \ensuremath{0.058}$  &            \ensuremath{0.0000} &              \ensuremath{0.01}  &   \ensuremath{\mathbf{0.3460}} &     \ensuremath{\mathbf{6.35}}  &           \ensuremath{-0.0050} &             \ensuremath{-0.08} \\
\midrule
\multicolumn{9}{l}{$\Lambda$CDM + Neutrinos + He} \\[2mm]
      $\Omega_b h^2$ & $      0.0226$ & $         \ensuremath{0.022602} \pm          \ensuremath{0.000061}$  &   \ensuremath{2\times 10^{-6}} &              \ensuremath{0.04}  &            \ensuremath{0.0001} &              \ensuremath{1.94}  &   \ensuremath{2\times 10^{-6}} &              \ensuremath{0.03} \\
      $\Omega_c h^2$ & $       0.112$ & $           \ensuremath{0.1121} \pm            \ensuremath{0.0014}$  &            \ensuremath{0.0001} &              \ensuremath{0.09}  &            \ensuremath{0.0007} &              \ensuremath{0.50}  &            \ensuremath{0.0001} &              \ensuremath{0.06} \\
               $H_0$ & $          70$ & $            \ensuremath{70.04} \pm              \ensuremath{0.54}$  &            \ensuremath{0.0400} &              \ensuremath{0.07}  &            \ensuremath{1.1000} &              \ensuremath{1.94}  &            \ensuremath{0.0100} &              \ensuremath{0.01} \\
              $\tau$ & $        0.09$ & $           \ensuremath{0.0900} \pm            \ensuremath{0.0022}$  &            \ensuremath{0.0000} &              \ensuremath{0.02}  &            \ensuremath{0.0001} &              \ensuremath{0.04}  &            \ensuremath{0.0001} &              \ensuremath{0.04} \\
               $n_s$ & $        0.96$ & $           \ensuremath{0.9601} \pm            \ensuremath{0.0028}$  &            \ensuremath{0.0001} &              \ensuremath{0.03}  &            \ensuremath{0.0032} &              \ensuremath{1.11}  &           \ensuremath{-0.0000} &             \ensuremath{-0.01} \\
 $\log(10^{10} A_s)$ & $      3.0445$ & $           \ensuremath{3.0448} \pm            \ensuremath{0.0051}$  &            \ensuremath{0.0003} &              \ensuremath{0.05}  &            \ensuremath{0.0016} &              \ensuremath{0.31}  &            \ensuremath{0.0003} &              \ensuremath{0.07} \\
             $N_\nu$ & $       3.046$ & $            \ensuremath{3.054} \pm             \ensuremath{0.087}$  &            \ensuremath{0.0080} &              \ensuremath{0.09}  &            \ensuremath{0.0890} &              \ensuremath{1.01}  &            \ensuremath{0.0040} &              \ensuremath{0.04} \\
               $Y_p$ & $        0.24$ & $           \ensuremath{0.2396} \pm            \ensuremath{0.0053}$  &           \ensuremath{-0.0004} &             \ensuremath{-0.07}  &   \ensuremath{\mathbf{0.0206}} &     \ensuremath{\mathbf{3.99}}  &           \ensuremath{-0.0005} &             \ensuremath{-0.09} \\
\midrule
\multicolumn{9}{l}{$\Lambda$CDM + Running} \\[2mm]
      $\Omega_b h^2$ & $      0.0226$ & $         \ensuremath{0.022594} \pm          \ensuremath{0.000043}$  &  \ensuremath{-6\times 10^{-6}} &             \ensuremath{-0.15}  &  \ensuremath{\mathbf{-0.0002}} &    \ensuremath{\mathbf{-4.05}}  &  \ensuremath{-2\times 10^{-6}} &             \ensuremath{-0.05} \\
      $\Omega_c h^2$ & $       0.112$ & $          \ensuremath{0.11203} \pm           \ensuremath{0.00055}$  &   \ensuremath{3\times 10^{-5}} &              \ensuremath{0.06}  &  \ensuremath{-7\times 10^{-5}} &             \ensuremath{-0.13}  &   \ensuremath{4\times 10^{-5}} &              \ensuremath{0.07} \\
               $H_0$ & $          70$ & $            \ensuremath{69.99} \pm              \ensuremath{0.23}$  &           \ensuremath{-0.0100} &             \ensuremath{-0.06}  &           \ensuremath{-0.1000} &             \ensuremath{-0.42}  &           \ensuremath{-0.0100} &             \ensuremath{-0.05} \\
              $\tau$ & $        0.09$ & $           \ensuremath{0.0901} \pm            \ensuremath{0.0022}$  &            \ensuremath{0.0001} &              \ensuremath{0.04}  &            \ensuremath{0.0006} &              \ensuremath{0.28}  &            \ensuremath{0.0001} &              \ensuremath{0.03} \\
               $n_s$ & $        0.96$ & $           \ensuremath{0.9598} \pm            \ensuremath{0.0017}$  &           \ensuremath{-0.0002} &             \ensuremath{-0.10}  &  \ensuremath{\mathbf{-0.0118}} &    \ensuremath{\mathbf{-6.98}}  &           \ensuremath{-0.0000} &             \ensuremath{-0.00} \\
 $\log(10^{10} A_s)$ & $      3.0445$ & $           \ensuremath{3.0449} \pm            \ensuremath{0.0047}$  &            \ensuremath{0.0004} &              \ensuremath{0.08}  &           \ensuremath{-0.0000} &             \ensuremath{-0.00}  &            \ensuremath{0.0004} &              \ensuremath{0.09} \\
    $n_\mathrm{run}$ & $           0$ & $          \ensuremath{-0.0001} \pm            \ensuremath{0.0027}$  &           \ensuremath{-0.0001} &             \ensuremath{-0.05}  &  \ensuremath{\mathbf{-0.0088}} &    \ensuremath{\mathbf{-3.41}}  &           \ensuremath{-0.0001} &             \ensuremath{-0.03} \\
\midrule
\multicolumn{9}{l}{$\Lambda$CDM (CV limited up to $l_\text{max} = 3000$)} \\[2mm]
      $\Omega_b h^2$ & $      0.0226$ & $         \ensuremath{0.022600} \pm          \ensuremath{0.000019}$  &           \ensuremath{-0.0000} &             \ensuremath{-0.03}  &  \ensuremath{\mathbf{-0.0002}} &   \ensuremath{\mathbf{-10.32}}  &   \ensuremath{3\times 10^{-6}} &              \ensuremath{0.18} \\
      $\Omega_c h^2$ & $       0.112$ & $          \ensuremath{0.11202} \pm           \ensuremath{0.00047}$  &   \ensuremath{2\times 10^{-5}} &              \ensuremath{0.03}  &           \ensuremath{-0.0002} &             \ensuremath{-0.38}  &   \ensuremath{4\times 10^{-5}} &              \ensuremath{0.09} \\
               $H_0$ & $          70$ & $            \ensuremath{69.99} \pm              \ensuremath{0.19}$  &           \ensuremath{-0.0100} &             \ensuremath{-0.03}  &           \ensuremath{-0.1000} &             \ensuremath{-0.52}  &           \ensuremath{-0.0100} &             \ensuremath{-0.06} \\
              $\tau$ & $        0.09$ & $           \ensuremath{0.0901} \pm            \ensuremath{0.0020}$  &            \ensuremath{0.0001} &              \ensuremath{0.07}  &           \ensuremath{-0.0021} &             \ensuremath{-1.18}  &            \ensuremath{0.0000} &              \ensuremath{0.01} \\
               $n_s$ & $        0.96$ & $           \ensuremath{0.9600} \pm            \ensuremath{0.0014}$  &            \ensuremath{0.0000} &              \ensuremath{0.03}  &  \ensuremath{\mathbf{-0.0175}} &   \ensuremath{\mathbf{-12.66}}  &            \ensuremath{0.0001} &              \ensuremath{0.08} \\
 $\log(10^{10} A_s)$ & $      3.0445$ & $           \ensuremath{3.0448} \pm            \ensuremath{0.0042}$  &            \ensuremath{0.0003} &              \ensuremath{0.08}  &           \ensuremath{-0.0060} &             \ensuremath{-1.62}  &            \ensuremath{0.0003} &              \ensuremath{0.06} \\
\bottomrule
\end{tabular}